\documentclass[11pt]{article}

\usepackage[preprint]{acl}

\usepackage{times}
\usepackage{latexsym}

\usepackage[T1]{fontenc}

\usepackage[utf8]{inputenc}

\usepackage{microtype}

\usepackage{inconsolata}

\usepackage{graphicx}
\usepackage{svg}
\usepackage{enumitem}
\usepackage{multirow}
\usepackage{tabularx}
\usepackage{makecell}
\usepackage{booktabs}
\usepackage{pifont}
\usepackage{xcolor}
\usepackage{colortbl}
\usepackage{tikz}
\usepackage{mathbbol}
\usepackage{relsize}

\usepackage{caption}
\usepackage{graphicx}
\usepackage{array}
\usepackage{pifont}            
\usepackage{amsmath}

\newcommand{\medbot}[1]{\textit{MeDxAgent}}
\newcommand{\dataset}[1]{\textit{MeDxBench}}

\definecolor{heroSage}{HTML}{B5CDA8}      
\definecolor{warnTerra}{HTML}{EAD2C2}     
\definecolor{oracleWarm}{HTML}{F2EDE2}    
\definecolor{labelGray}{HTML}{666666}
\definecolor{nameTint}{HTML}{F7F2E8}
\definecolor{labelGray}{HTML}{666666}
\newcolumntype{P}[1]{>{\raggedright\arraybackslash}p{#1}}
\definecolor{checkGreen}{HTML}{2E7D32}   
\definecolor{xmarkGray}{HTML}{C5C5C5}    

\newcommand{\cmarkc}{\textcolor{checkGreen}{\ding{51}}}  
\definecolor{xmarkRed}{HTML}{D88B8B}
\newcommand{\xmarkc}{\textcolor{xmarkRed}{\ding{55}}}

\usepackage{hyperref}        
\usepackage[breakable,skins]{tcolorbox}

\newtcolorbox{HighlighterBox}[1]{
  enhanced,
  breakable=true,
  colback=yellow!8,
  colframe=yellow!50!black,
  fonttitle=\bfseries,
  fontupper=\sffamily\small,           
  title=#1,
  pad at break*=2mm,
  boxrule=0.5pt,
  bottomrule=0.5pt,
  toprule=0.5pt,
  bottomrule at break=0.5pt,           
  toprule at break=0.5pt,              
}

\newtcolorbox{DialogueExampleBox}[1]{
  enhanced,
  breakable=true,
  colback=cyan!5,
  colframe=cyan!55!black,
  fonttitle=\bfseries,
  fontupper=\sffamily\small,
  title=#1,
  pad at break*=2mm,
  boxrule=0.5pt,
  bottomrule=0.5pt,
  toprule=0.5pt,
  bottomrule at break=0.5pt,
  toprule at break=0.5pt,
}

\title{\medbot{}: Multi-Agent Consultation for Interactive Medical Diagnosis}

\author{
 \textbf{Akshat Sanghvi\textsuperscript{1}},
 \textbf{Naren Akash\textsuperscript{2}},
 \textbf{Raza Imam\textsuperscript{3}},
 \textbf{Amit Sharma\textsuperscript{1}},
 \textbf{Mohit Jain\textsuperscript{1}}
\\
\\
 \textsuperscript{1}Microsoft Research India,\\
 \textsuperscript{2}Institute of AI and Innovation, AIG Hospitals,\\
 \textsuperscript{3}Mohamed bin Zayed University of Artificial Intelligence
\\
 \small{
   \textbf{Correspondence:} \href{mailto:t-asanghvi@microsoft.com}{t-asanghvi@microsoft.com}, \href{mailto:mohja@microsoft.com}{mohja@microsoft.com}
 }
}

\begin{document}
\maketitle

\begin{abstract}

Large language models (LLMs) are increasingly used for health-related decision support.
Yet most evaluations treat diagnosis as a single-shot task with complete information provided upfront, often as a multiple-choice selection.
This diverges from clinical practice, where diagnosis is interactive and open-ended, involving sequential hypothesis refinement through targeted questioning.
We address this gap.
We build \dataset{}, a large-scale benchmark of $4{,}421$ clinical cases across 20 specialties.
We further propose \medbot{}, a multi-agent consultation system for interactive diagnosis, and systematically study its prompt-, flow- and agent-level design choices.
\medbot{} achieves a 10.3\% accuracy gain over the baseline on \dataset{}, closing 52.3\% of the gap to a full-information oracle.
We find that specific design choices: collecting demographics first, passing summarized dialogue for diagnosis, and feeding candidate diagnoses for targeted questioning, improve accuracy, mirroring how physicians reason, though their effect emerges fully only in combination.
Code and dataset will be released upon publication.

\end{abstract}

\begin{figure*}[t]
    \centering
    \includegraphics[width=0.95\textwidth]{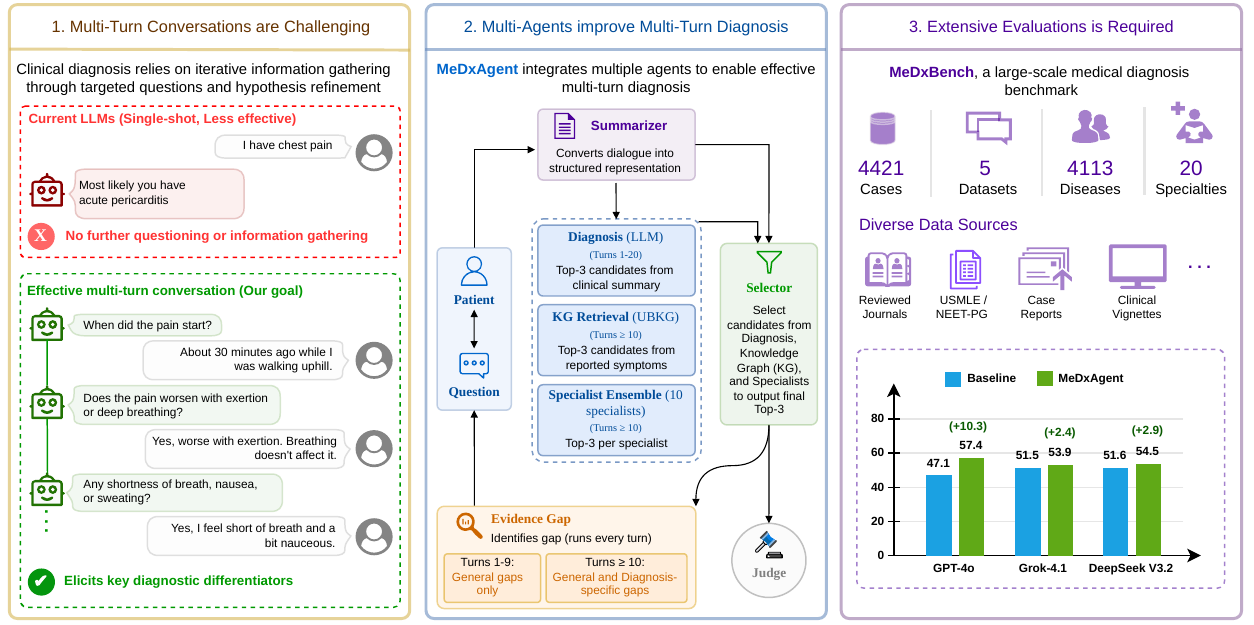}
    \caption{Limitations in existing systems. Our proposed \medbot{} agentic system and \dataset{} benchmark.}
    \label{fig:teaser}
\end{figure*}

\section{Introduction}
\label{sec:introduction}

LLMs are increasingly used for real-world decision support, with healthcare emerging as one of the most prominent application domains~\cite{copilot-usage-mai,copilot-health-usage-mai}.
A substantial fraction of LLM usage involves health-related queries, with users relying on them to interpret symptoms and determine next steps.
However, LLMs largely operate in a single-shot setting, producing diagnoses from incomplete symptom descriptions without iterative information gathering.
In contrast, clinical diagnosis is inherently interactive:
physicians ask targeted questions, iteratively refine hypotheses, and perform differential diagnosis.

Despite this, most benchmarks for medical diagnosis evaluate LLMs in static task formulation.
Models are provided complete case descriptions, often exam-style questions such as USMLE, and evaluated on their ability to predict the correct diagnosis~\cite{medqa, medpalm}.
Although LLMs perform strongly on such benchmarks, in some cases matching or exceeding physicians~\cite{gpt5mmmed, medpalm2}, the task lacks ecological validity.

Recent work~\cite{amie, vivabench, medagentsim} has begun exploring multi-turn doctor-patient interactions using LLM agents for disease diagnosis.
However, several limitations remain.
\textbf{First}, there is no standardized diagnostic benchmark that combines diverse clinical cases with reliable diagnoses.
Existing datasets are fragmented across formats (e.g., clinical vignettes, EHRs) and specialties (e.g., dermatology, cardiology), and mix diagnoses with unrelated tasks.
\textbf{Second}, existing agentic approaches remain limited.
Open systems often use simplistic paradigm of a single “doctor” agent that iteratively questions the patient until it determines enough information has been gathered for diagnosis~\cite{vivabench, mediq, aie}, whereas the sophisticated multi-agent systems are closed-sourced and cannot be reproduced or compared~\cite{amie, mai-dxo}.
\textbf{Third}, several setups leak diagnostic information and thereby weaken the task:
For example, some initialize the doctor with part of the case (e.g., by passing the first sentence of the patient's case description) \cite{mediq},
some reduce diagnosis to multiple-choice selection \cite{mediq, jia2025ddo, rose-etal-2025-meddxagent}, 
some allow the patient agent to generate synthetic information beyond the case description using the ground truth diagnosis~\cite{mai-dxo, dynamicare, rose-etal-2025-meddxagent, wang2025medkgi}, and some convert case descriptions into structured fields that occasionally contain information revealing diagnosis~\cite{medagentsim}.
These choices can artificially inflate accuracy.
\textbf{Fourth}, prior work typically reports results from a single system configuration, 
leaving it unclear which design choices are responsible for observed gains.

We study this design space systematically. 
We first construct \dataset{}, a large-scale diagnosis benchmark of 4{,}421 clinical cases collected from five public sources, spanning 20 specialties and ranging from common to rare diseases.
Our simulated patient agent answers only from the case description, preventing diagnosis information leaks during conversation. 
We then propose \medbot{}, a multi-agent consultation system for interactive diagnosis. We evaluate its \textit{prompt-}, \textit{flow-}, and \textit{agent-level} design choices independently and in combination, treating each choice as a hypothesis about the capabilities required for effective diagnosis.

\medbot{} improves diagnostic accuracy by 10.3\% over the baseline on \dataset{}, and recovers 52.3\% of the gap between the baseline and a full-information oracle.
Our analysis reveals that three capabilities prove particularly important: collecting patient demographics first, summarizing doctor-patient dialogue into a clinical problem representation, and feeding diagnostic reasoning to guide subsequent questioning. 
Further, we reveal two broader findings.
First, the effectiveness of an intervention depends strongly on \textit{when} it is applied.
For example, we introduce differential questioning, which directs the agent to ask questions that distinguish among competing diagnoses. Applying this strategy from the second versus the tenth turn produces an 18.1\% variation in diagnostic accuracy.
Second, components that hurt performance in isolation can become \textit{beneficial in combination}. Three of the four agents reduce accuracy individually, but the full system integrating all four outperforms every partial configuration. A standard incremental ablation study would have incorrectly rejected these components.



\textbf{Contributions.}

\begin{itemize}[nosep,noitemsep]
\item \textbf{A large-scale medical diagnosis benchmark, \dataset{}.} 
We standardize 4{,}421 clinical cases from five public datasets, spanning 20 specialties and a range of disease prevalence.
\item \textbf{A multi-agent consultation system, \medbot{}.} Our system gathers information through targeted questions, summarizes clinical evidence, maintains competing hypotheses, and commits to a diagnosis (similar to a physician). It recovers 52.3\% of the gap between a baseline agent and a full-information oracle, and the gains transfer across model families.
\item \textbf{A systematic study of design choices for interactive diagnosis.} We evaluate prompt-, flow-, and agent-based variations independently and in combination, showing that both intervention timing and component interactions critically shape performance. 
\end{itemize}


\section{Related Work}
\label{sec:related_work}

\begin{table}[t]
\centering
\scriptsize
\caption{Comparison of interactive diagnostic systems.}
\label{tab:relatedwork_comparison}
\vspace{2pt}

\setlength{\tabcolsep}{3.5pt}
\renewcommand{\arraystretch}{1.15}

\begin{tabular}{lccccc}
\toprule
\textbf{Method} &
\makecell{\textbf{Multi-}\\\textbf{Agent}} &
\makecell{\textbf{Open-ended}\\\textbf{Diagnosis}} &
\makecell{\textbf{No}\\\textbf{Leakage}} &
\makecell{\textbf{Open}\\\textbf{Data, Code}} &
\makecell{\textbf{\#}\\\textbf{Cases}} \\
\midrule

MediQ         & \xmarkc & \xmarkc & \cmarkc & \cmarkc & $1{,}412$ \\
MEDDxAgent    & \cmarkc & \xmarkc & \xmarkc & \cmarkc & $300$ \\
DDO           & \cmarkc & \xmarkc & \xmarkc & \xmarkc & $1{,}684$ \\
DynamiCare    & \cmarkc & \cmarkc & \xmarkc & \xmarkc & $500$ \\
MAI-DxO       & \cmarkc & \cmarkc & \xmarkc & \xmarkc & $304$ \\
MedKGI        & \cmarkc & \cmarkc & \xmarkc & \xmarkc & N/A \\
AMIE          & \xmarkc & \cmarkc & \cmarkc & \xmarkc & $149$ \\
AIHospital    & \xmarkc & \cmarkc & \xmarkc & \cmarkc & $506$ \\
MedAgentSim   & \cmarkc & \cmarkc & \xmarkc & \cmarkc & $622$ \\
VivaBench     & \xmarkc & \cmarkc & \cmarkc & \cmarkc & $990$ \\
\rowcolor{heroSage}
\textbf{\medbot{}} &
\cmarkc &
\cmarkc &
\cmarkc &
\cmarkc &
\textbf{$4{,}421$} \\

\bottomrule
\end{tabular}
\end{table}

\textbf{Static Diagnostic Benchmarks.}
Most medical LLM benchmarks evaluate diagnosis in a static setting, where the full case is provided upfront and the model predicts a diagnosis in a single turn, as in MultiMedQA~\cite{medpalm}, HealthBench~\cite{healthbench}, DiagnosisArena~\cite{diagnosis_arena} and MedHELM~\cite{medhelm}.
LLMs now match or exceed expert-level performance on several such benchmarks~\cite{gpt5mmmed, medpalm2}.
Multi-agent approaches such as MDAgents~\cite{mdagents}, MDTeamGPT ~\cite{chen2025mdteamgpt}, Agent Hospital~\cite{agenthospital}, and MedAgents~\cite{medagents} assign specialized roles to different agents, but still operate on fully observed cases.
All these approaches sidesteps the iterative information-gathering central to clinical diagnosis.

\noindent \textbf{Interactive Diagnostic Systems.}
Recent work has begun exploring simulated doctor-patient interactions for diagnosis.
Some approaches formulate the task as disease selection rather than open-ended diagnosis prediction.
MediQ~\cite{mediq} studies iterative questioning to identify the correct diagnosis from multiple choices; DDO~\cite{jia2025ddo} and MEDDxAgent~\cite{rose-etal-2025-meddxagent} frame diagnosis as classification over a limited disease set.
Other works focus on benchmarking LLMs by prompting them to act as clinicians conducting follow-up questioning, rather than designing multi-agent systems. CRAFT-MD~\cite{craftmd} introduces a conversational evaluation framework spanning multiple specialties; 
VivaBench~\cite{vivabench} studies cognitive failure modes in LLM diagnosis, and identifies anchoring bias.
Most recently, agentic approaches have been developed for open-ended conversational diagnosis.
AMIE~\cite{amie}, optimized for diagnostic reasoning and conversations, demonstrated physician-level performance on 149 simulated primary-care consultations with trained human actors; MAI-DxO~\cite{mai-dxo} orchestrated a panel of physician-persona agents to diagnose 304 complex NEJM cases at physician-level accuracy.
However, both systems are closed-source and their evaluation protocols are not reproducible.
Several open systems have also been proposed, such as AgentClinic~\cite{agentclinic}, AI Hospital~\cite{aihospital} and MedAgentSim~\cite{medagentsim}.
However, these works evaluate either a single doctor agent configuration or have relatively small benchmarks, limiting systematic analysis of design choices (Table~\ref{tab:relatedwork_comparison}).
In contrast, we systematically study how prompt design, interaction flow, and agent composition affect multi-turn diagnostic performance across a large benchmark.




\section{\dataset{} Benchmark Dataset}
\label{sec:datasets}

\dataset{} is a large-scale benchmark for medical diagnosis containing 4,421 cases spanning 20 clinical specialties (Table~\ref{table:dataset}). 
Each case includes a unique identifier, case description, source dataset, specialty, and ground-truth diagnosis label(s).


\textbf{Sources.}
We define a diagnostic case as a patient case description paired with a ground-truth diagnosis.
We build \dataset{} by aggregating and standardizing cases from five public datasets that meet this criterion: CRAFT-MD~\cite{craftmd} ($200$ cases), DiagnosisArena~\cite{diagnosis_arena} ($897$), MedMCQA~\cite{medmcqa} ($847$), MedQA~\cite{medqa} ($1{,}498$), and PubMed cases curated by VivaBench~\cite{vivabench} ($988$).

\textbf{Statistics.}
The dataset spans diverse case formats, difficulties, and specialties.
Case formats include exam-style questions (MedQA, MedMCQA), free-text clinical vignettes (CRAFT-MD), and peer-reviewed case reports (DiagnosisArena, PubMed).
Its $4{,}113$ unique disease labels comprise $655$ common (high prevalence, $>1$ in $100$ globally), $1{,}098$ uncommon (moderate prevalence, $1$ in $100$ to $1$ in $2{,}000$) and $2{,}360$ rare ($<1$ in $2{,}000$) conditions.
It spans $20$ specialties, with the largest proportions from Infectious Diseases ($11.1\%$), Neurology ($9.7\%$) and Emergency Medicine ($8.9\%$).
Case descriptions range from $212$ to $9{,}848$ characters (mean $1{,}785$ $\pm$ $1{,}955$), and $546$ cases have more than one ground-truth diagnosis labels.

\textbf{Construction.} 
From the five datasets, we keep only diagnostic cases with sufficient clinical detail.
We remove cases focused on medical knowledge, treatment recommendations, or other non-diagnostic tasks. We keep only descriptions longer than 200 characters.
Because our work studies text-only diagnosis, we exclude cases requiring image/video interpretation, while retaining those containing textual imaging reports.
We also remove duplicate cases appearing across datasets.

To prevent \textbf{diagnosis leakage}, we redact explicit mentions of the ground-truth diagnosis from each case description.
We manually verify that redaction does not remove clinically relevant evidence.




\section{Simulation Framework}
\label{sec:method}
Our framework simulates multi-turn doctor-patient conversations using three agents:
(i) a Patient agent that answers from a case description,
(ii) a Doctor agent that conducts the consultation and
(iii) a Judge agent that evaluates the final diagnosis.
All agents are implemented through prompt engineering; full prompts are provided in Appendix~\ref{appendix_prompts}.
The simplest configuration serves as our \textit{baseline}.
Its Doctor agent comprise of two components: a Question agent that interacts with the patient and a Diagnosis agent that predicts candidate diagnoses after each turn. Our full system, \medbot{}, adds a summarizer, specialist ensemble, knowledge-graph module, and evidence-gap finder (Section~\ref{sec:variants}).

At initialization, the Patient receives the case description and the Judge receives the ground-truth diagnosis.
The Doctor begins with a standard opening question (``What brings you in today?''), to which the Patient responds with the primary concern.
At each subsequent turn, the Doctor asks a question conditioned on the conversation history, and the Patient responds strictly from the case description, returning ``I don't know'' if the requested information is not present.
After each turn, the Doctor outputs three candidate diagnoses, each with a short rationale and a confidence score. The interaction continues until either a diagnosis reaches a 95\% confidence threshold, selected on a 300-case validation split, or a 20-turn limit is reached (following \cite{mediq}).
In the latter case, the highest-confidence diagnosis from the last turn is selected.
The final prediction is then evaluated by the Judge.
All experiments use temperature $=0$.

\textbf{Patient Agent}.
The Patient answers strictly from the case description, without fabricating or inferring unstated details, responding with ``I don't know'' otherwise.
This prevents diagnosis leakage, 
since the Patient has access only to the case description and not the ground-truth label.
Cases are typically written retrospectively (e.g., as clinical notes), so we instruct the agent to respond as if currently experiencing the condition, in the first person, and in plain language.

\textbf{Doctor Agent.}
The baseline Doctor has two components, both operating at each turn.
\textbf{(1) Question Agent} generates the next query and a brief rationale, from the conversation history.
We instruct it to ask simple, atomic questions covering key clinical dimensions (including demographics, symptoms, medical and family history, physical examination findings, and test results), to avoid repeating questions, to prioritize diagnostically informative questions, and to shift its line of inquiry when the Patient answers "I don't know."
\textbf{(2) Diagnosis Agent} produces three candidate diagnoses from the conversation history, each with a short justification and self-reported confidence score.
The agent is not conditioned on its own prior predictions; Section~\ref{sec:variants} studies that choice directly.

\textbf{Judge Agent}.
The Judge uses LLM-based semantic matching to
determine whether the predicted and ground-truth diagnosis refer to the same condition.
A prediction is considered correct if it is (i) an exact match, (ii) a recognized synonym or paraphrase (e.g., high BP vs. hypertension), or (iii) a more specific subtype of the ground-truth (e.g., malignant melanoma for melanoma).
Predictions more general than the ground-truth are considered incorrect.
For multi-label cases, a match with any ground-truth label counts as correct, since multiple diagnoses can explain the same presentation.

\section{\medbot{} System}
\label{sec:variants}
We study how design choices affect multi-turn diagnosis.
We treat each design choice as a hypothesis and group them by where they intervene:
(1) \textit{Prompt-based} choices modify the instructions provided to the Question agent, (2) \textit{Flow-based} choices change the interaction flow, and (3) \textit{Agent-based} choices introduce additional agents for summarization, reasoning, or knowledge grounding.
We evaluate each choice incrementally(Table~\ref{tab:ablation_gpt4o}).
The Patient and Judge agents remain fixed across all variants, ensuring that performance differences arise only from the design choice under study.
We use clinician analogies to motivate these design choices, not to claim physician-like reasoning.

\subsection{Prompt-Based Variations}
We evaluate two prompt modifications to the baseline, as well as their combination.

\textbf{Turn Awareness.}
We provide the Question agent with the remaining question budget and turn-specific guidance: early turns (1-5) focus on chief complaints and high-yield screening questions, middle turns (6-15) gather objective evidence and differentiate symptoms, and late turns (16-20) prioritize the most discriminative questions. Because early questions shape the utility of later follow-ups, question ordering compounds over the consultation. We test whether turn budget awareness helps the agent plan and allocate its questions better.

\textbf{Demographics First.}
We require the Question agent to collect age and sex in the first turn.
This mirrors clinical practice, where demographic information helps anchor the differential diagnosis
\cite{bordage}. 
In the baseline, the agent is encouraged, but not required, to ask about demographics.
We test whether fixing demographics at the start orients the diagnostic process by ruling out age- and sex-dependent implausible candidates.

\textbf{Combined Prompt.}
We combine both modifications; apply turn awareness at every step and require demographic collection in the first turn.

\subsection{Flow-Based Variations}
Building on the best prompt variant (Table \ref{tab:ablation_gpt4o}), \textit{Combined Prompt}, we study flow-based modifications.

\textbf{No Early Stopping.}
The baseline ends a consultation once the Diagnosis agent reports $\geq95\%$ confidence.
Here, we remove this and run all consultations for the full 20 turns, and select the highest-confidence diagnosis from the final turn.
We test whether gathering more evidence before committing improves diagnostic accuracy.

\textbf{Differential Questioning.}
In the baseline, the next question is generated independently of the current diagnosis predictions.
Here, the Question agent additionally receives the candidate diagnoses, confidence scores, and supporting reasoning, and is instructed to ask questions that differentiate competing hypotheses. This is motivated by hypothetico-deductive reasoning, in which physicians form hypotheses early and gather evidence to confirm or reject them~\cite{elstein}.
This strategy has a known failure mode: conditioning too early on weak hypotheses can lead to \textit{anchoring}~\cite{graber, vivabench}. 
We therefore activate differential questioning only after a fixed turn threshold, testing switch points at turns 2, 5, and 10.
Activating at turn 2 enables it almost from the start, since the first turn collects demographics; switching at later turns allow more general evidence to accumulate before conditioning future questions on diagnostic hypotheses. 

\textbf{Combined Flow.}
We combine no early stopping with differential questioning activated at turn 10.


\subsection{Agent-Based Variations}
Building on the best Flow variant, \textit{Differential Questioning (Turn 10)}, we introduce four agents.

\textbf{Summarizer.}
Instead of passing the raw conversation to the Diagnosis agent, a Summarizer converts it into a structured clinical representation.
A physician distills an encounter into a concise problem representation that organizes findings and activates relevant diagnosis~\cite{bowen}.
Additionally, LLMs are trained primarily on case-style medical text (e.g., clinical notes, USMLE-style vignettes) rather than conversational transcripts, making summarized representations better aligned with their training data.
We test two formats:
(a) a free-form paragraph-style clinical narrative, and
(b) a structured SOAP-style schema~\cite{universal-abstraction} with fields such as \texttt{symptoms\_present}, \texttt{medical\_history}, and \texttt{test\_results}.


\textbf{Specialist Ensemble.}
We use an ensemble of ten physician agents: nine specialists (Cardiology, Pulmonology, Gastroenterology, Neurology, Endocrinology, Nephrology, Rheumatology, Haematology, and Dermatology) and one general physician.
Each agent reviews the conversation and predicts its top three diagnoses with confidence scores and reasoning, producing a diverse candidate pool.
A separate Selector-SE agent, designed as a senior physician, selects the final top three diagnosis by weighting specialists according to their relevance to the current case rather than by majority voting. This follows multidisciplinary clinical practice, where diverse specialists contribute to a single case. We test whether such diversity surfaces hypotheses missed by a single generalist agent.

\textbf{Knowledge Graph.}
Structured medical knowledge sources can surface relevant candidates and improve the factual grounding of predictions~\cite{drknows}.
We therefore ground candidate diagnoses generation in the Unified Biomedical Knowledge Graph (UBKG)~\cite{ubkg}.
At each turn, we query UBKG using symptoms revealed by the Patient and aggregate the diseases it returns. Diseases are scored based on the number of matched symptoms, and the top three are passed to the Selector-KG agent.
As UBKG does not incorporate clinical evidence such as test results, the base Diagnosis agent also generates its own top three predictions using the conversation history.
The Selector-KG receives both sets and selects a final top three based on consistency with the conversation.
We test whether knowledge graph grounding contributes complementary hypotheses that the LLM overlooks.

\textbf{Evidence Gap.}
Clinicians gather information selectively, directing questions toward the evidence that would most change the differential diagnosis~\cite{bowen}. 
The Evidence Gap agent follows this principle, identifying information still needed to support or reject a diagnosis.
It identifies:
(i) \textit{diagnosis-specific gaps} that help differentiate current candidate diagnoses, and
(ii) \textit{general gaps} covering broadly informative clinical details (such as vitals, lab results).
The agent outputs up to three gaps per type, expressed as short, actionable statements, ranked by estimated diagnostic utility.
Before turn 10, when candidate diagnoses remain unused by the Question agent, it reports only general gaps using the dialog history. From turn 10, when Differential Questioning is active, it also receives the top three candidates and reports both gap types.
The Question agent is prompted to prioritize these gaps when forming the next query.
We test whether explicitly modeling missing evidence improves information gathering.

\textbf{Combined Agents: \medbot{}.}
It integrates all four agents (Figure~\ref{fig:teaser}).
The Summarizer runs at every turn, converting dialogue into a structured clinical representation.
During early turns (1-9), the Question agent gathers information broadly, guided only by general evidence gaps.
From turn 10, the system enters a differential diagnosis phase. The Knowledge Graph retrieves symptom-matched diseases, the Specialist Ensemble proposes additional specialty-specific hypotheses, and the Selector consolidates these candidates into a final top three.
The Evidence Gap agent now reports both general and diagnosis-specific gaps, enabling the Question agent to ask question based on current hypotheses and missing evidence.

\section{Results}
\label{sec:results}

%
%
%

\begin{table*}[t]
\centering
\caption{Diagnostic accuracy of \medbot{} variants 
using GPT-4o. Bold marks the best result within each section. Note: Results are averaged over three runs, with variance below 0.5\% across runs.
}
\label{tab:ablation_gpt4o}
\scriptsize
\resizebox{\textwidth}{!}{%
\begin{tabular}{llcccccc}
\toprule
\textbf{Choice} & \textbf{Variant} & \textbf{Avg. Acc.} & \textbf{CRAFT-MD (200)} & \textbf{Diag. Arena (897)} & \textbf{MedMCQA (847)} & \textbf{MedQA (1489)} & \textbf{PubMed (988)} \\
\midrule
& Baseline & 47.1 & 52.5 & 18.8 & 57.3 & 45.9 & 60.9 \\
\midrule
\textit{\color{labelGray}Prompt} & Turn Awareness            & 48.6 & 56.5 & 21.1 & 59.2 & 45.5 & 61.0 \\
& Demographics First        & 49.8 & 54.0 & 20.4 & 61.0 & 51.4 & 62.4 \\
& \textbf{Combined Prompt}           & \textbf{51.7} & \textbf{57.0} & \textbf{22.5} & \textbf{63.9} & \textbf{51.9} & \textbf{63.3} \\
\midrule
{\textit{\color{labelGray}Flow}} & No Early Stopping         & 51.6 & 54.5 & \textbf{24.9} & 62.6 & 50.1 & \textbf{66.1} \\
& Differential Q (Turn 2)   & 34.7 & 35.0 & 12.3 & 47.7 & 36.9 & 41.6 \\
& Differential Q (Turn 5)   & 50.4 & 59.0 & 18.8 & 63.6 & 52.6 & 57.7 \\
& \textbf{Differential Q (Turn 10)}  & \textbf{52.8} & 60.5 & 23.5 & \textbf{64.5} & 54.2 & 61.1 \\
& Combined Flow             & 52.8 & \textbf{61.0} & 20.9 & 63.8 & 54.1 & 64.4 \\
\midrule
{\textit{\color{labelGray}Agent}} & Summarizer: Paragraph     & 54.9 & 61.0 & \textbf{27.4} & 64.6 & 54.8 & 66.6 \\
& Summarizer: Structured    & 54.6 & 61.0 & 27.0 & 63.2 & 55.1 & 67.0 \\
& Specialist Ensemble      & 51.1 & 54.5 & 24.2 & 63.9 & 52.3 & 60.6 \\
& Knowledge Graph           & 51.7 & 57.0 & 21.6 & 62.9 & 54.3 & 62.6 \\
& Evidence Gap              & 50.6 & 59.5 & 23.2 & 61.0 & 48.8 & 60.5 \\
\rowcolor{heroSage}
& \textbf{\medbot}          & \textbf{57.4} & \textbf{66.5} & \textbf{27.4} & \textbf{69.8} & \textbf{56.9} & \textbf{66.6} \\
\midrule
\rowcolor{oracleWarm}
& \textit{Oracle (Full Information)} & \textit{66.8} & \textit{65.5} & \textit{31.7} & \textit{76.3} & \textit{68.6} & \textit{92.0} \\
\bottomrule
\end{tabular}%
}
\end{table*}

Using GPT-4o as the underlying model, \medbot{} achieves 57.4\% average diagnostic accuracy on \dataset{}, 10.3 points above the baseline (47.1\%) (Table~\ref{tab:ablation_gpt4o}).
Performance improves monotonically across the prompt-, flow-, and agent-level additions.
Gains hold across all five constituent datasets, although absolute accuracy varies: from 27.4\% on DiagnosisArena to 66.6\% on PubMed.
The full information Oracle, which receives the complete case upfront, reaches only 66.8\%, indicating that the task is far from solved even with perfect information.
Against this ceiling, \medbot{} closes 52.3\% of the baseline-to-Oracle gap.
The remaining 9.4-point gap is likely due to these reasons: degradation of LLM performance in multi-turn settings~\cite{LLMs_lost_multiturn}, incomplete information due to frequent ``I don't know'' responses from the Patient agent, and possible training-data contamination suggested by PubMed's near-ceiling Oracle accuracy (92.0\%).

\textbf{Different variants contribute differently.}
Combined Prompt reaches 51.7\%, improving over the baseline by 4.6 points through two changes to the Question agent: collecting demographics in the first turn and conditioning on the remaining turn budget.
Combined Flow reaches 52.8\%, driven primarily by differential questioning. Treating differential questioning as a timed intervention rather than an always-on capability reveals strong sensitivity.
Activating it at turn 2 reduces accuracy to 34.7\%, 12.4 points below the baseline. Delaying activation to turn 5 recovers most of the loss (50.4\%) and turn 10 achieves 52.8\%, yielding an 18.1-point swing from changing only the switch point.
Removing early stopping does not improve performance over Combined Prompt (51.6\% vs.\ 51.7\%), which suggests that the Diagnosis agent's confidence-based stopping criterion is well-calibrated: consultations end early (mean 16.3 turns) once sufficient evidence is gathered.

\textbf{The agent layer shows strong compositional effects.} The agent layer provides the largest single gain, from 52.8\% at Differential Questioning (Turn 10) to 57.4\% with all four agents (+4.6 points). 
Individually, only the Summarizer improves performance, in both formats (+2.1 paragraph, +1.8 structured). The other three agents each reduce accuracy by 1-2 points when added in isolation. 
Even though the sum of the four individual gains is negative, the full system outperforms every partial configuration. 
Leave-one-out ablations further confirm this dependence: removing any agent from \medbot{} reduces accuracy below the full system (Table~\ref{tab:removal_ablation}), indicating that no agent is redundant. The largest accuracy drop occurs when removing the (Summarizer) agent, which contributed the most.

\textbf{Cross-model transfer.}
The design choices developed on GPT-4o transfer to other model families, Grok-4.1 and DeepSeek-V3.2, without re-tuning prompts or thresholds (Table~\ref{tab:model_variation}). 
Absolute gains are smaller on Grok-4.1 (51.5 to 53.8) and DeepSeek-V3.2 (51.5 to 54.5) than on GPT-4o (47.1 to 57.4), but each layer consistently improves over the baseline across models, and \medbot{} remains the strongest variant, which shows the design generalizes beyond the model it was developed on.
The larger gains on GPT-4o may stem from greater headroom for improvement, since it begins further from its Oracle ceiling than Grok-4.1 and DeepSeek-V3.2.
\begin{table}[t]
\centering
\scriptsize

\caption{Diagnostic accuracy of \medbot{} variants using different LLMs. D1: CRAFT-MD, D2: Diagnosis Arena, D3: MedMCQA, D4: MedQA, D5: PubMed.}
\label{tab:model_variation}
\begin{tabular}{clccccc}
\toprule
\textbf{Model} & \textbf{Variant} & \textbf{D1} & \textbf{D2} & \textbf{D3} & \textbf{D4} & \textbf{D5} \\
\midrule
{\textit{\color{labelGray}GPT-4o}} & Baseline     
& 52.5 & 18.8 & 57.3 & 45.9  & 60.9 \\
& Combined Prompt        
& 57.0 & 22.5 & 63.9 & 51.9  & 63.3 \\
& Differential Q        
& 60.5 & 23.5 & 64.5 & 54.2  & 61.1 \\

\rowcolor{heroSage}
& \textbf{\medbot{}}
& \textbf{66.5}
& \textbf{27.4}
& \textbf{69.9}
& \textbf{56.9}
& \textbf{66.6} \\

\midrule

\rowcolor{oracleWarm}
& \textit{Full Info.}
& \textit{65.5}
& \textit{31.7}
& \textit{76.3}
& \textit{68.6}
& \textit{92.0} \\

\midrule

\textit{\color{labelGray}Grok-4.1} & Baseline     
& 53.5 & 26.6 & 60.7 & 47.4 & 69.4 \\
& Combined Prompt        
& 55.0 & 26.9 & 60.6 & 48.4 & \textbf{70.6} \\
& Differential Q        
& 58.0 & 26.8 & 62.6 & 50.0 & 67.3 \\

\rowcolor{heroSage}
& \textbf{\medbot{}}
& \textbf{62.0}
& \textbf{28.8}
& \textbf{63.8}
& \textbf{51.4}
& 63.8 \\

\midrule

\rowcolor{oracleWarm}
& \textit{Full Info.}
& \textit{60.5}
& \textit{35.5}
& \textit{74.5}
& \textit{62.2}
& \textit{92.6} \\

\midrule

\multirow{2}{*}{\makecell[c]{\textit{\color{labelGray}DeepSeek}\\ \textit{\color{labelGray}V3.2}}}
& Baseline       
& 56.0 & 26.3 & 61.6 & 48.2 & \textbf{65.8} \\
& Combined Prompt        
& 55.5 & 25.3 & 63.2 & 48.9 & 64.4 \\
& Differential Q        
& 61.5 & \textbf{27.9} & \textbf{67.2} & 51.0 & 65.7 \\

\rowcolor{heroSage}
& \textbf{\medbot{}}
& \textbf{65.0}
& 26.3
& 66.6
& \textbf{54.3}
& 60.5 \\

\midrule

\rowcolor{oracleWarm}
& \textit{Full Info.}
& \textit{73.0}
& \textit{36.2}
& \textit{78.0}
& \textit{68.0}
& \textit{91.5} \\

\bottomrule
\end{tabular}

\end{table}


%

\begin{table}[t]
\centering
\scriptsize
\caption{Comparison of \medbot{} with prior work.}
\label{tab:prior_work_comparison}
\begin{tabular}{lccc}
\multicolumn{4}{l}{\textbf{(a) \medbot{} vs.\ MedAgentSim} \textit{\color{labelGray}— GPT-4o backbone}} \\
\midrule
\textbf{Dataset (Size)}
& \textbf{MedAgentSim}
& \textbf{\medbot{}}
& \textbf{$\Delta$} \\
\midrule
NEJM (120)    & 27.50 & \textbf{41.67} & +14.17 \\
MIMIC (288)   & 75.30 & \textbf{76.04} & +0.74 \\
MedQA (214)   & 67.80 & \textbf{71.50} & +3.70 \\
\midrule
Avg.          & 56.87 & \textbf{63.07} & +6.20 \\
\bottomrule

\multicolumn{4}{l}{} \\
\multicolumn{4}{l}{\textbf{(b) \medbot{} vs.\ VivaBench} \textit{\color{labelGray}— PubMed benchmark}} \\
\midrule
\textbf{Model}
& \textbf{VivaBench}
& \textbf{\medbot{}}
& \textbf{$\Delta$} \\
\midrule
GPT-4o         & 23.1 & \textbf{43.3} & +20.2 \\
o4-mini        & 32.0 & \textbf{37.7} & +5.7 \\
Gemini-2.5-Pro & 35.0 & \textbf{43.5} & +8.5 \\
\bottomrule
\end{tabular}

\end{table}









\begin{table}[t]
\centering
\scriptsize
\caption{Leave-one-out ablations on \medbot{}.}
\label{tab:removal_ablation}

\begin{tabular}{cccccc}
\toprule

\textbf{\medbot{}} &
\makecell{w/o \\ Summarizer} &
\makecell{w/o Specialist \\ Ensemble} &
\makecell{w/o KG} &
\makecell{w/o Evidence \\ Gap} \\

\midrule

\textbf{57.4\%} &
54.4\% &
55.7\% &
56.0\% &
55.5\% \\

\bottomrule
\end{tabular}%
\end{table}

\textbf{Comparison with prior work.}
We compare \medbot{} against recent diagnostic systems,
MedAgentSim~\cite{medagentsim} and VivaBench~\cite{vivabench}, on their respective evaluation datasets and protocols.
\medbot{} outperforms MedAgentSim on all three benchmarks (Table~\ref{tab:prior_work_comparison}a), improving average accuracy from 56.9\% to 63.1\% (+6.2 points), with the largest gain on NEJM (+14.2 points).
\medbot{} also surpasses VivaBench on the Pubmed benchmark across models (Table~\ref{tab:prior_work_comparison}b), with gains of +20.2 points with GPT-4o and +8.5 points with Gemini-2.5-Pro.
Note: Scores on Pubmed differ from Table~\ref{tab:ablation_gpt4o} because Vivabench uses a different evaluation setup.
These improvements indicate that \medbot{}'s design choices transfer across datasets and model families.

\section{Discussion}
\label{sec:discussion}

\textbf{What worked and why.} Three design choices account for most of the gains, and each aligns with established principles of clinical reasoning. First, collecting demographics first orient the differential diagnosis because age and sex strongly shape disease priors.
For instance, in 39 pediatric cases, demographics were the key differentiating factor in diagnosis: the baseline defaulted to adult presentations and missed age-specific conditions, whereas the demographics-first variant correctly identified them.
This matches prior reports of demographic biases in medical LLMs~\cite{diversitymedqa} and shows a single-line prompt intervention can help mitigate them in an agentic system. Second, Differential Questioning helps, but only after sufficiently strong hypotheses have formed.
This mirrors hypothetico-deductive pattern of clinical reasoning, in which physicians form hypotheses and gather evidence to confirm or rule them out~\cite{elstein}, and its associated failure mode, anchoring, where weak early hypotheses bias subsequent reasoning~\cite{graber}. Our experiments show that the same instruction and interaction flow can improve or reduce performance depending solely on when it is activated. Third, the Summarizer improves accuracy 
by changing how evidence is represented rather than what evidence is available. Clinicians refer to this as problem representation, and prior work links higher-quality representations to more accurate diagnosis~\cite{bowen}. Also, summarization may help because structured clinical summaries 
more closely resemble the case-style medical text on which LLMs are predominantly trained.

\textbf{Why three agents fail in isolation.}
The remaining three agents lower accuracy individually, each for a different reason.
The Specialist Ensemble and Knowledge Graph (KG) variants are prone to premature termination (Table~\ref{tab:round_analysis_variants}). In the Specialist Ensemble, repeated predictions across specialists inflate aggregate confidence during ensembling. In the KG variant, the Selector-KG overweights KG-derived candidates. These causes early stopping before sufficient evidence is gathered, reducing accuracy.
The Evidence Gap agent appears to fail due to confirmation bias. Although it extracts more information, as reflected by a sharp reduction in ``I don't know'' answers relative to the baseline (Figure~\ref{fig:idk_dist}), it tends to direct questions toward confirming the leading hypothesis. When that hypothesis is incorrect, the system gathers increasingly detailed evidence for the wrong disease, lowering performance. Despite this, each agent contributes positively in \medbot{}. Leave-one-out ablations (Table \ref{tab:removal_ablation}) show that removing any agent from the full system lowers accuracy.
We do not claim that the combined system fully resolves these issues; however, we provide empirical evidence for why the combined setup leads to improved accuracy.

\textbf{Components work in combination.}
The strong compositional effects arise because the modules are designed to operate jointly.
The Summarizer produces a structured representation, which helps the Knowledge Graph and Specialist Ensemble generate `diverse' diagnostic candidates. These candidates enable Evidence Gap to identify missing evidence more reliably, which in turn helps the Question agent to ask more targeted questions and reduce "I don't know" responses.
As discussed, individually, the Knowledge Graph and Specialist Ensemble tend to be overconfident, leading to premature termination. When combined, however, their predictions are balanced, reflected in lower early stopping rates (Table \ref{tab:round_analysis_variants}).
Similarly, the Evidence Gap agent alone is prone to confirmation bias, but benefits from the more diverse candidate set produced by the combined system, reducing anchoring effects.
Overall, each agent in isolation perturbs a system not built to absorb it, often reducing accuracy. Together, the agents become complementary, and removing any one lowers accuracy. Prior multi-agent medical systems typically report aggregate gains over single-agent baselines without isolating individual agent contributions~\cite{mai-dxo, mdagents, medagents}, leaving such compositional effects unexplored. We highlight this phenomenon empirically, though
we do not claim it generalizes to all multi-agent systems.


\textbf{Real-world deployment.}
Despite promising performance, \medbot{} should be viewed strictly as a decision-support tool for clinicians, not a replacement for medical professionals. It is not suitable for autonomous use by non-medical personnel.
Incorrect or overconfident diagnoses could delay treatment, trigger unnecessary interventions, or cause serious patient harm, particularly in high-risk clinical settings.
Moreover, conversational diagnostic systems may inherit biases (e.g., demography-related) and reasoning failures from their underlying LLMs.
In practice, such systems may be most appropriate for assisting physicians with differential diagnosis, surfacing alternative hypotheses, or double-checking clinical decisions.

\section{Conclusion}
\label{sec:conclusion}

We present \dataset{}, a benchmark of $4{,}421$ clinical cases across 20 specialities for open-ended interactive diagnosis and \medbot{}, a multi-agent system built from a systematic study of prompt-, flow- and agent-level design choices. 
Beyond these, our contribution is methodological: we isolate the impact of individual design decisions and reveal interaction effects.
\section{Limitations}
First, our evaluation is limited to text-only diagnosis, while real-world clinical diagnosis often depends on multimodal evidence such as imaging.
Second, the Patient agent answers strictly from the case description, which prevents diagnosis leakage, but reduces the natural variability and ecological validity of real patient interactions, as patients rarely respond with repeated "I don't know" responses. Finally, although we systematically study several design choices, a broader sweep over prompts, interaction flows, and agent configurations would strengthen the empirical conclusions; though such sweeps are computationally expensive in multi-turn agentic settings. Future work can address these limitations and evaluate whether the design choices that help in simulation also transfer to supervised clinical use, with multimodal inputs, and learned questioning policies beyond prompt engineering.
\bibliography{references}

%
%

\appendix


\section*{Appendix}


\section{Additional Results}
\label{appendix_additional_results}


\subsection{Accuracy by Specialty}

\begin{figure*}
    \centering
    \includegraphics[width=1\linewidth]{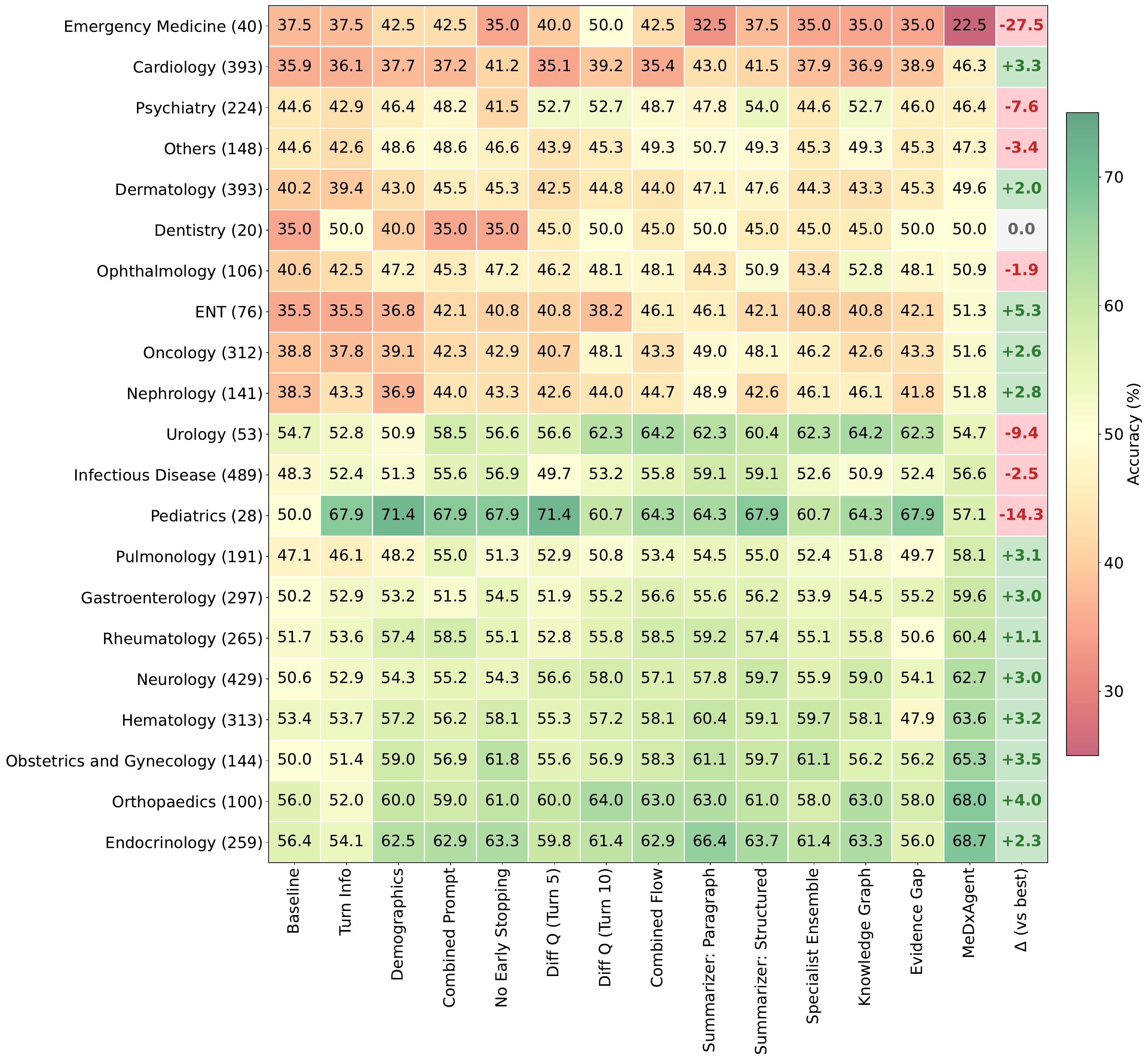}
    \caption{Diagnostic accuracy by specialty across variants. \medbot{} (rightmost) is the strongest variant in 15 of 20 specialties.}
    \label{fig:specialty_accuracy_dist}
\end{figure*}

\begin{figure*}
    \centering
    \includegraphics[width=1\linewidth]{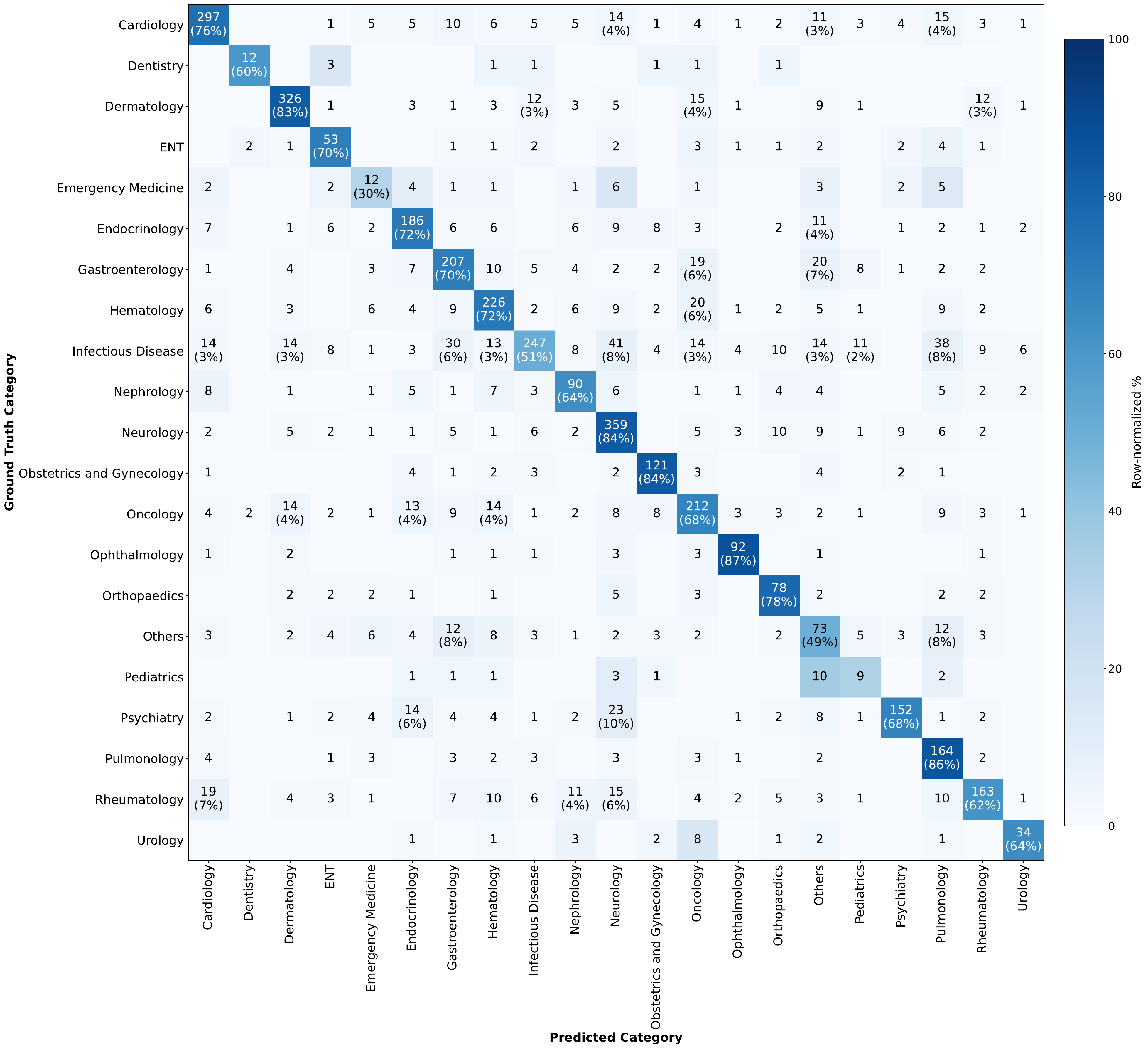}
    \caption{Specialty wise confusion matrix for \medbot{}, Ground Truth (GT) vs Predicted Specialty. Percentage values are row-normalized, and are shown only for cells with greater than 10 cases.}
    \label{fig:confusion_matrix}
\end{figure*}

Figure~\ref{fig:specialty_accuracy_dist} reports diagnostic accuracy across 20 specialties for all variants, and Figure~\ref{fig:confusion_matrix} shows the confusion matrix between ground-truth and predicted specialties.
\medbot{} is the best performing variant in 15 of 20 specialties, with its highest accuracies in Endocrinology (68.7\%), Orthopaedics (68.0\%), and Obstetrics \& Gynaecology (65.3\%). It improves over the baseline in all specialties except Emergency Medicine, where its predictions are frequently shifted towards Neurology and Pulmonology.

The confusion matrix yields a specialty-match accuracy (i.e., the predicted diagnosis belongs to the correct specialty) of 70.4\%. 
Most specialty confusions are clinically adjacent.
Ophthalmology (86.8\%), Obstetrics \& Gynaecology (84.0\%), Neurology (83.7\%), and Dermatology (83.0\%) achieve the highest specialty-match accuracy, indicating that these specialties have distinctive clinical presentations that our system rarely confuses with other domains.
In contrast, Rheumatology errors are scattered across Cardiology, Neurology and Nephrology, consistent with autoimmune diseases presenting through the organ systems they affect. In such cases, our system predicts an organ-specific diagnosis rather than the underlying systemic disorder.

\begin{figure*}
    \centering
    \includegraphics[width=1\linewidth]{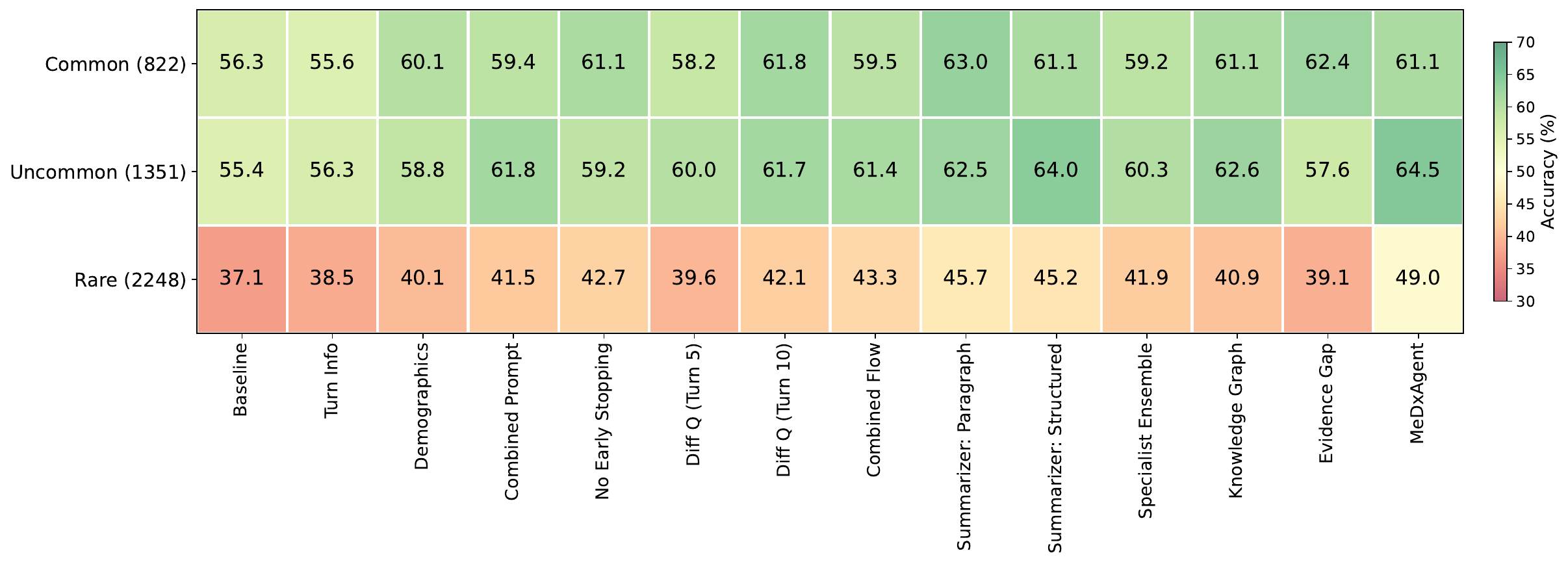}
    \caption{Diagnostic accuracy by disease prevalence. \medbot{}'s gain over the baseline increases from common to rare conditions (+4.8, +9.1, +11.9 points).}
    \label{fig:difficulty_heatmap}
\end{figure*}

\subsection{Accuracy by Disease Prevalence}

Figure~\ref{fig:difficulty_heatmap} reports accuracy across three disease prevalence tiers. For cases with multiple ground-truth diagnoses, evaluation uses the most prevalent diagnosis among the labels. \medbot{}'s gains over the baseline is inversely correlated with disease prevalence: +4.8 points on common diseases (56.3\% to 61.1\%), +9.1 points on uncommon diseases (55.4\% to 64.5\%), and +11.9 points on rare diseases (37.1\% to 49.0\%).
Among individual components, the Summarizer (paragraph format) performs best on rare cases (45.7\% against 42.1\% for the best flow variant), consistent with summarization helping most when findings are sparse and scattered across the conversation, as they often are in rare presentations.


\subsection{Top-k Accuracy}
Table \ref{tab:topk_comparison} reports top-1, top-2, and top-3 accuracies for \medbot{} and the full-information Oracle. \medbot{} reaches 68.0\% top-3 accuracy, compared to 77.7\% for the Oracle.
%
%
%

\begin{table*}[t]
\footnotesize
\centering
\caption{Overview of datasets aggregated into \dataset{}. \textbf{Format}: V = free-text vignette, CR = journal case report, EQ = exam-style questions. \textbf{Dx}: S = single, M = multiple ground-truth diagnoses per case.}
\label{table:dataset}
\begin{tabular}{>{\columncolor{nameTint}}P{2.3cm} P{5cm} r r c c P{2.5cm}}
\toprule
\textbf{Dataset} & \textbf{Details} & \textbf{Size} & \textbf{Filtered} & \textbf{Format} & \textbf{Dx} & \textbf{Major Specialties} \\
\midrule
\textbf{\mbox{CRAFT-MD}} & Online vignettes \& expert- curated cases & 200 & 200 & V & S & Derm, ID, Peds, Cardio, Neuro \\
\addlinespace[3pt]
\textbf{DiagnosisArena} & Peer-reviewed journal case reports (e.g., Cell, JAMA) & 915 & 897 & CR & S & Derm, Onco, ID, Cardio, Neuro \\
\addlinespace[3pt]
\textbf{MedMCQA} & MCQs from Indian medical entrance exams (AIIMS and NEET-PG)& 194k & 847 & EQ & S & Peds, Neuro, Onco, ID, EM \\
\addlinespace[3pt]
\textbf{MedQA} & USMLE-style board exams; MedReason reasoning chains & 12k & 1{,}489 & EQ & S & Peds, ID, EM, Neuro, Gastro \\
\addlinespace[3pt]
\textbf{PubMed} & Physician-curated PubMed case reports & 988 & 988 & CR & M & ID, Neuro, Gastro, Cardio, Rheum \\
\bottomrule
\end{tabular}%
\end{table*}
%
%

\begin{table*}[t]
\centering
\footnotesize
\caption{Round analysis across all variants. Early Stop (\%) is computed as $1 - \left(\frac{\text{Total Rounds}}{\text{\#Cases} \times \text{Max Rounds}}\right)$, representing the fraction of interaction rounds saved relative to the maximum budget of 20 rounds per case. ``IDK Rounds'' and ``IDK Cases'' denote the proportion of rounds and cases, respectively, in which the patient lacked the requested information and responded with "I don't know".}
\label{tab:round_analysis_variants}
\begin{tabular}{lccccc}
\toprule
\textbf{Variant} & \textbf{Accuracy} & \textbf{Early Stop (\%)} & \textbf{Total Rounds} & \textbf{IDK Rounds (\%)} & \textbf{IDK Cases (\%)} \\
\midrule
Baseline                    & 47.1 & 18.4 & 72,119 & 62.3 & 89.8 \\
Turn Awareness              & 48.6 & 18.6 & 71,979 & 60.6 & 89.8 \\
Demographics First          & 49.8 & 18.1 & 72,396 & 59.9 & 88.1 \\
\textbf{Combined Prompt}             & 51.7 & 19.0 & 71,610 & 59.2 & 87.4 \\
No Early Stopping           & 51.6 & 0.0 & 88,420 & 62.5 & 100.0 \\
Differential Q (Turn 2)     & 34.7 & 12.9 & 76,982 & 68.7 & 94.7 \\
Differential Q (Turn 5)     & 50.4 & 19.4 & 71,282 & 60.9 & 87.7 \\
\textbf{Differential Q (Turn 10)}    & 52.8 & 19.9 & 70,826 & 60.0 & 87.2 \\
Combined Flow               & 52.8 & 0.0 & 88,420 & 61.8 & 100.0 \\
Summarizer: Paragraph                & 55.2 & 4.2  & 84,726 & 61.6 & 98.0 \\
Summarizer: Structured                 & 54.6 & 8.1  & 81,293 & 61.0 & 96.6 \\
Specialist Ensemble                 & 51.1 & 23.3 & 67,787 & 61.0 & \textbf{83.4} \\
Knowledge Graph             & 51.7 & 17.6 & 72,891 & 60.8 & 88.3 \\
Evidence Gap                & 50.6 & 4.8  & 84,154 & \textbf{57.4} & 99.0 \\
\rowcolor{heroSage}
\textbf{\medbot{}} & \textbf{57.4} & \textbf{4.2} & \textbf{84,739} & 58.0 & 99.2 \\
\bottomrule
\end{tabular}
\end{table*}
%
%

\begin{table*}[t]
\centering
\footnotesize
\caption{Top-$3$ diagnostic accuracy (\%) of \medbot{} and Full Information setting}
\label{tab:topk_comparison}
\begin{tabular}{llcccccc}
\toprule
\textbf{$k$} & \textbf{Method} & \textbf{Average} & \textbf{CraftMD} & \textbf{DiagnosisArena} & \textbf{MedMCQA} & \textbf{MedQA} & \textbf{PubMed} \\
\midrule

\multirow{2}{*}{Top-1}
& \medbot{}          & 57.4 & 66.5 & 27.4 & 69.8 & 56.9 & 66.6 \\
& Full Information   & 66.8 & 65.5 & 31.7 & 76.3 & 68.6 & 92.0 \\
\midrule

\multirow{2}{*}{Top-2}
& \medbot{}          & 65.7 & 73.0 & 36.6 & 77.9 & 65.7 & 75.1 \\
& Full Information   & 75.6 & 76.5 & 41.3 & 85.9 & 78.0 & 96.2 \\
\midrule

\multirow{2}{*}{Top-3}
& \medbot{}          & 68.0 & 74.5 & 39.8 & 80.0 & 68.6 & 77.3 \\
& Full Information   & 77.7 & 77.5 & 44.9 & 88.6 & 80.8 & 96.8 \\
\bottomrule
\end{tabular}
\end{table*}

\subsection{Accuracy of the Patient Agent}
We evaluate the Patient agent using an LLM-as-a-judge over all responses generated by \medbot{}, totaling 84{,}739 question-answer pairs across 4,421 cases. The Patient agent achieves an overall response accuracy of 97.2\%.
Among the 2{,}387 inaccurate responses (2.8\%), we identify four error types: \textit{withholding} (1{,}513 responses; 63.4\% of errors), where the Patient omits information present in the case description, typically failing to mention a detail when asked a broad question; \textit{fabrication} (457; 19.1\%), where the Patient introduces clinical details not present in the case; \textit{contradiction} (368; 15.4\%), where the response conflicts with information in the case; and \textit{oversharing} (48; 2.0\%), where the Patient provides information not solicited by the doctor's question. 
Overall, these results suggest that the Patient agent remains highly faithful to the underlying case descriptions.

\subsection{Round-Level Conversation Analysis}
Table \ref{tab:round_analysis_variants} reports per-variant accuracy, early stopping rate, total rounds, and the rate of ``I don't know'' (IDK) responses from the Patient agent. The early stopping rate is defined as the fraction of the 20-turn budget left unused, computed as $1 - \left(\frac{\text{Total Rounds}}{\text{\#Cases} \times \text{Max Rounds}}\right)$.
Across variants, roughly 60\% of rounds and more than 80\% of cases contain at least one IDK response, reflecting how often requested clinical details are absent from the case description.
The Specialist Ensemble exhibits the highest early stopping rate (23.3\%) and the lowest IDK case rate (83.4\%), suggesting that its inflated diagnostic confidence ends consultations earlier, leaving fewer turns to ask unanswerable questions.

Across variants, accuracy correlates with the rate of informative answers. The correlation between accuracy and the fraction of ''I don't know'' (IDK) rounds is $-0.79$, while the correlation with total rounds ($0.27$) and with the raw count of IDK rounds ($-0.07$) is weak. Thus variants whose questions more often draw substantive answers tend to be more accurate.

\begin{figure*}
    \centering
    \includegraphics[width=1\linewidth]{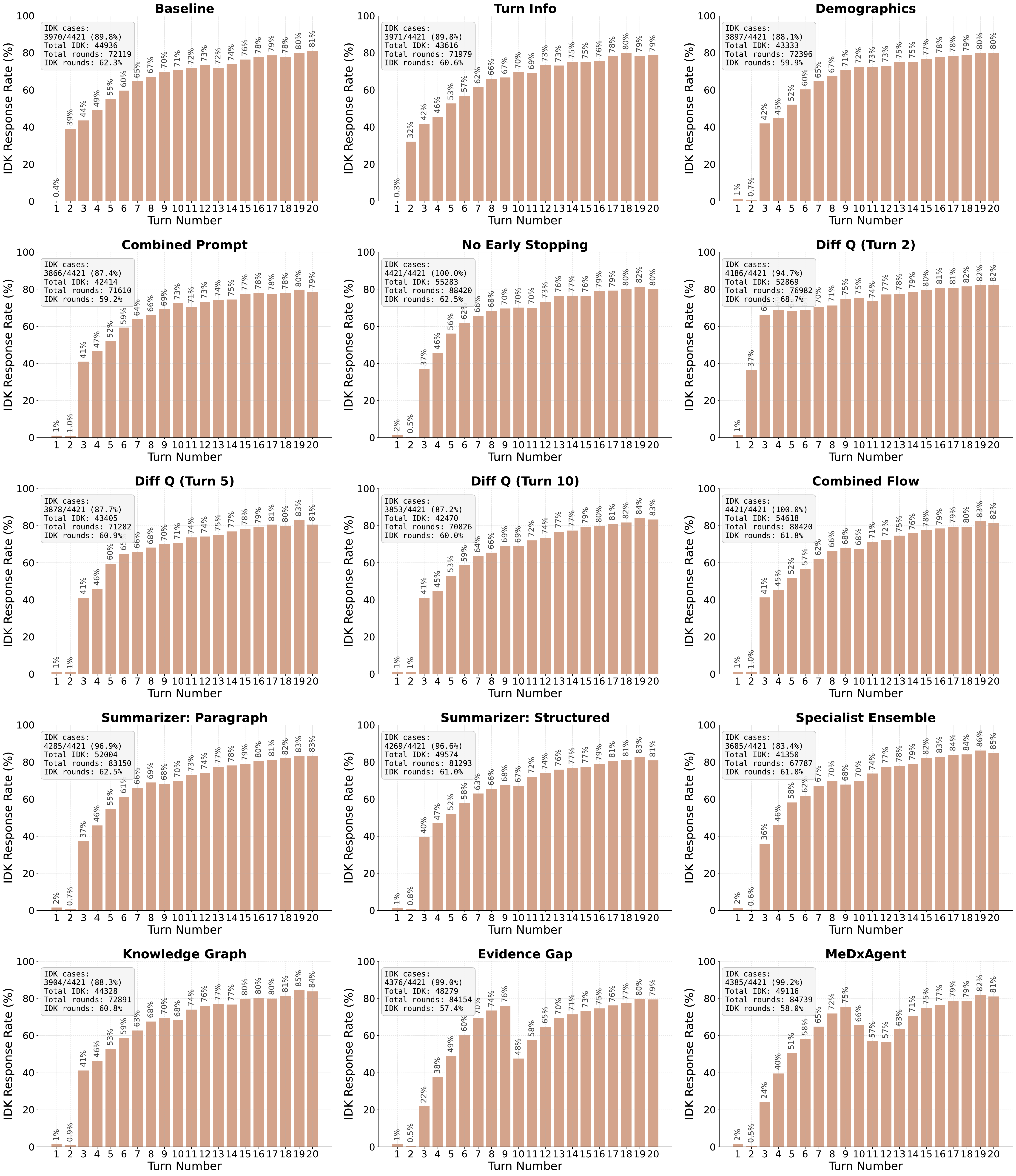}
    \caption{Per-turn rate of "I don't know" patient responses by variant. The Evidence Gap variant and \medbot{} show a distinctive dip near turn 10 when gap-directed questioning activates.}
    \label{fig:idk_dist}
\end{figure*}

Figure \ref{fig:idk_dist} shows the per-turn IDK rate of each variant. Most variants show a characteristic increasing IDK rate across turns, as early turns extract the readily available demographics and chief complaints, while later turns probe more specific clinical details that are often missing in the case description. The Evidence Gap variant shows a distinct pattern.
Its gap-directed questioning lowers the IDK rate around turn 10 by targeting diagnostically relevant missing information. The same pattern appears in \medbot{}. Section \ref{sec:discussion} uses this behavior to explain the observed compositional effects.

\begin{figure*}
    \centering
    \includegraphics[width=1\linewidth]{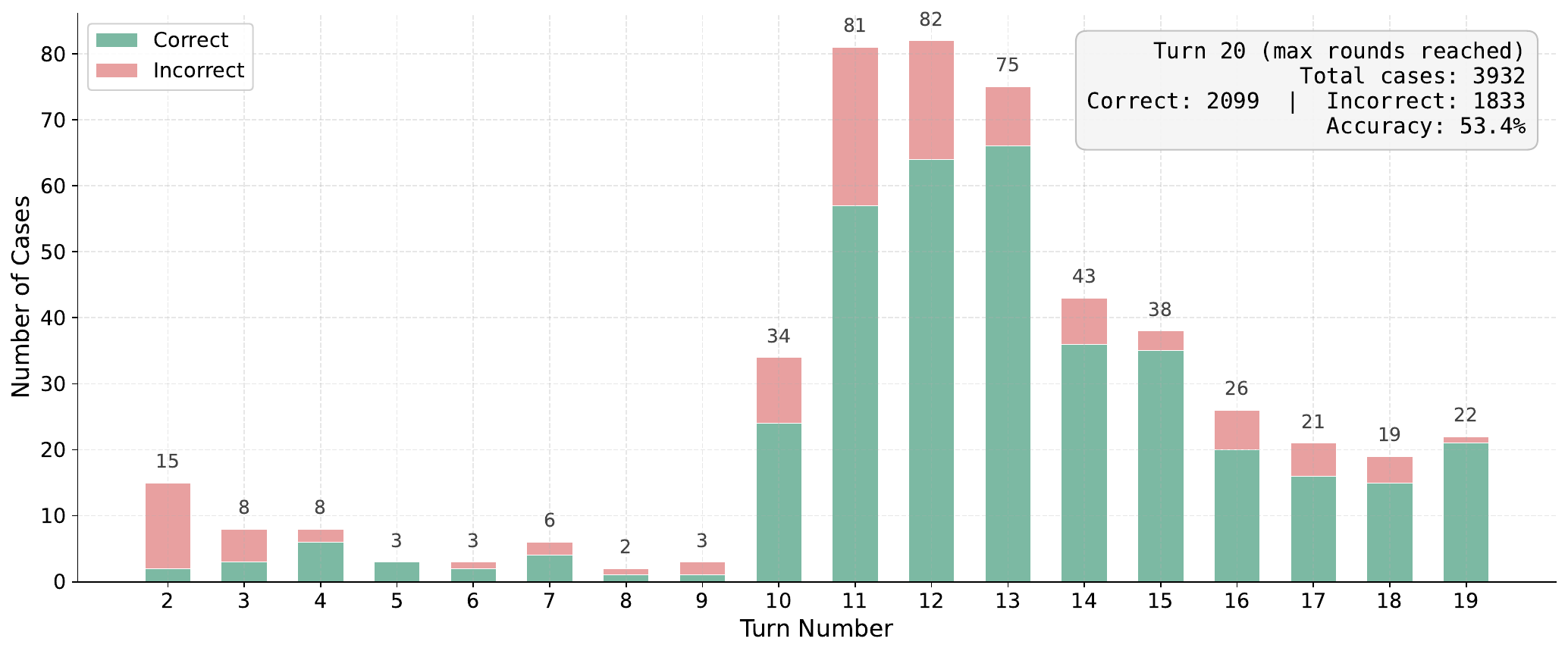}
    \caption{Distribution of cases terminating at each dialogue turn for \medbot{} variant. Cases terminating at turn 20 are shown separately in the inset box. }
    \label{fig:stop_dist}
\end{figure*}

Figure \ref{fig:stop_dist} shows when consultations terminate. For \medbot{}, termination rates increase sharply from turn 10 onward, when the full multi-agent pipeline (Specialist Ensemble, Knowledge Graph, and Evidence Gap agents) becomes active. This indicates that these components substantially boost diagnostic confidence, often pushing prediction confidence above the 95\% stopping threshold. Even so, 3,932 out of 4,421 (88.9\%) still run the full 20 turns, suggesting that most cases in the benchmark are not resolved with high-confidence within the available 20-turn budget.

\section{Implementation Details}
\label{appendix_implementation}
The simulation framework agents, the Patient and Judge agents, were fixed to GPT-4o across all experiments to ensure a consistent evaluation environment.
The remaining agents were instantiated using the model under evaluation: GPT-4o (\texttt{gpt-4o-2024-11-20}), DeepSeek-V3.2 (\texttt{DeepSeek-V3.2}), or Grok-4-1 (\texttt{grok-4-1-fast-non-reasoning}).
All LLM calls were made via the Azure OpenAI API with $temperature=0$ and $max\_tokens=4096$. Agent outputs were constrained using OpenAI's structured output mode (\texttt{response\_format}) with Pydantic schemas, enabling server-side enforcement of JSON schema compliance.

Dataset details are summarized in Table~\ref{table:dataset}, including the source dataset, number of original and filtered cases, case format (free-text vignette, case report, or exam-style question), diagnostic annotation setting (single vs.\ multiple ground-truth diagnoses), and major clinical specialties represented.

\section{Prompts}
\label{appendix_prompts}

This appendix provides an overview of all prompts used in our experiments, along with the different prompt versions and their activation conditions across system configurations. We organize them by role: the patient simulator (Section~\ref{appendix_prompts_patient}), the doctor's question agents (Section~\ref{appendix_prompts_question}), the doctor's diagnosis agents (Section~\ref{appendix_prompts_diagnosis}), the summarizer agents (Section~\ref{appendix_prompts_summarizer}), the evidence gap agent (Section~\ref{appendix_prompts_evidence}), and the judge agent (Section~\ref{appendix_prompts_judge}).

\subsection{Patient Agent}
\label{appendix_prompts_patient}

The patient agent is initialized with a case description and answers the doctor's questions strictly from that description, returning ``I don't know'' when information is unavailable. This prevents leakage of implicit disease knowledge.

\begin{HighlighterBox}{Patient Agent Prompt}

\textbf{Role.}
You are a patient in a medical consultation. You are provided with your complete medical history, including symptoms, family history, physical examination findings, and laboratory or imaging results. You are experiencing this case in the present.

\textbf{Instruction.}
Answer the doctor's question using only the information explicitly provided in the patient history.

\textbf{Rules.}
\begin{itemize}
    \item Answer only from the provided patient history.
    \item If the information is not available, respond with: ``I don't know.''
    \item Do not ask questions or provide explanations.
    \item Do not infer or invent information.
    \item Answer succinctly in first person, using simple non-technical language.
    \item Provide only the information asked for.
\end{itemize}

\textbf{Input.}
\begin{itemize}
    \item Patient history: \{patient\_history\}
    \item Doctor's question: \{question\}
\end{itemize}

\textbf{Output format.}
\begin{verbatim}
{
  "answer": "patient answer"
}
\end{verbatim}

\end{HighlighterBox} \vspace{0.5cm}

\subsection{Doctor: Question Agents}
\label{appendix_prompts_question}

Depending on the variant and question-asking objective, we use four question-agent versions in the experiments: Question Agent (\textit{QA}), Differential Question Agent (\textit{DQA}), Evidence-Guided Question Agent (\textit{EG-QA}) and Evidence-Guided Differential Question Agent (\textit{EG-DQA}). Below we summarize their activation across variants:
\begin{itemize}
    \item Baseline: \textit{QA} all turns
    \item Prompt-Based Variations: \textit{QA} all turns
    \item Flow-based Variations:
    \begin{itemize}
        \item No Early Stopping: \textit{QA} All turns
        \item Differential Q (turn $X$): \textit{QA} before turn $X$, \textit{DQA} after.
        \item Combined Flow: \textit{QA} before turn $10$, \textit{DQA} after
    \end{itemize}
    \item Agent-Based Variations
    \begin{itemize}
        \item Evidence Gap: \textit{EG-QA} before turn $10$, \textit{EG-DQA} after.
        \item \medbot: \textit{EG-QA} before turn $10$, \textit{EG-DQA} after.
        \item Rest: QA before turn $10$, \textit{DQA} after.
    \end{itemize}
\end{itemize}

\begin{HighlighterBox}{Question Agent Prompt}

\textbf{Role.}
You are an experienced physician conducting a diagnostic consultation.

\textbf{Instruction.}
Ask exactly one diagnostic question that maximizes information gain and helps narrow the diagnosis.

\textbf{Question scope.}
You may ask about demographics, current symptoms, symptom details, medical history, medications, family history, physical examination findings, vital signs, laboratory results, imaging findings, exposures, pregnancy, immunosuppression, or other relevant risk factors.

\textbf{Rules.}
\begin{itemize}
    \item Ask one clear, specific, atomic question.
    \item Do not repeat or rephrase any previously asked question.
    \item Avoid semantically similar questions about already-covered topics.
    \item Prefer high-yield questions that differentiate likely diagnoses.
    \item Ask for comprehensive test or examination results when relevant.
    \item If the patient repeatedly answers ``I don't know'' to a topic, move to a different topic.
    \item Do not ask to perform or order tests; ask whether results are already available.
    \item Avoid tunnel vision by periodically exploring different clinical domains.
\end{itemize}

\textbf{Round-aware strategy. (when enabled)}
\begin{itemize}
    \item Early rounds: establish chief complaint, demographics, vital signs, and broad high-yield information.
    \item Middle rounds: prioritize objective data and differentiating symptoms.
    \item Late rounds: ask the most discriminative questions between likely diagnoses.
    \item If only 1--3 rounds remain, ask only the highest-value confirmatory or exclusionary question.
\end{itemize}

\textbf{Input.}
\begin{itemize}
    \item Round status: \{round\_info\}
    \item Questions already asked: \{asked\_questions\}
    \item Interaction so far: \{dialog / summary\}
\end{itemize}

\textbf{Output format.}
\begin{verbatim}
{
  "question": "next diagnostic question"
}
\end{verbatim}

\end{HighlighterBox} \vspace{0.5cm}

\begin{HighlighterBox}{Differential Question Agent Prompt}

\textbf{Role.}
You are an experienced physician conducting a diagnostic consultation with access to the current differential diagnosis.

\textbf{Instruction.}
Ask exactly one new question that best distinguishes between the current top-3 diagnoses.

\textbf{Rules.}
\begin{itemize}
    \item Ask one clear, specific, atomic question.
    \item Do not repeat or rephrase any previously asked question.
    \item Focus on questions that shift confidence between competing diagnoses.
    \item Prefer findings that support one diagnosis while weakening alternatives.
    \item Prioritize objective evidence such as vital signs, laboratory tests, imaging, ECG, biopsies, cultures, and physical examination findings.
    \item Ask for comprehensive results when requesting tests.
    \item If only 1--3 rounds remain, ask the single most discriminative question between the top diagnoses.
\end{itemize}

\textbf{Input.}
\begin{itemize}
    \item Round status: \{round\_info\}
    \item Top-3 differential diagnoses: \{predictions\}
    \item Questions already asked: \{asked\_questions\}
    \item Interaction so far: \{dialog / summary\}
\end{itemize}

\textbf{Output format.}
\begin{verbatim}
{
  "question": "next diagnostic question"
}
\end{verbatim}

\end{HighlighterBox} \vspace{0.5cm}

\begin{HighlighterBox}{Evidence-Guided Question Agent Prompt}

\textbf{Role.}
You are an experienced physician using an evidence-gap report to guide information gathering.

\textbf{Instruction.}
Ask exactly one question that addresses the most important missing information identified in the evidence-gap report.

\textbf{Rules.}
\begin{itemize}
    \item Ask one clear, specific, atomic question.
    \item Do not repeat or rephrase any previously asked question.
    \item Use the evidence-gap report as the primary guide.
    \item Prioritize clinically important gaps that improve diagnosis.
    \item Ask for comprehensive results when requesting tests.
    \item If only 1--3 rounds remain, ask the single most important missing question.
    \item If more rounds remain, address evidence gaps in priority order.
\end{itemize}

\textbf{Input.}
\begin{itemize}
    \item Round status: \{round\_info\}
    \item Evidence-gap report: \{eg\_report\}
    \item Questions already asked: \{asked\_questions\}
    \item Interaction so far: \{dialog / summary\}
\end{itemize}

\textbf{Output format.}
\begin{verbatim}
{
  "question": "next diagnostic question"
}
\end{verbatim}

\end{HighlighterBox} \vspace{0.5cm}

\begin{HighlighterBox}{Evidence-Guided Differential Question Agent Prompt}

\textbf{Role.}
You are an experienced physician using both the current differential diagnosis and an evidence-gap report.

\textbf{Instruction.}
Ask exactly one new question that fills the most critical evidence gap and best distinguishes between the top-3 diagnoses.

\textbf{Rules.}
\begin{itemize}
    \item Ask one clear, specific, atomic question.
    \item  Do not repeat or rephrase any previously asked question.
    \item Use diagnosis-specific gaps to identify discriminative questions.
    \item Prefer questions that would substantially shift confidence between diagnoses.
    \item Ask for comprehensive test or examination results when relevant.
    \item If only 1--3 rounds remain, focus on gaps that confirm or exclude the leading diagnosis.
    \item If more rounds remain, address high-value evidence gaps systematically.
\end{itemize}

\textbf{Input.}
\begin{itemize}
    \item Round status: \{round\_info\}
    \item Top-3 differential diagnoses: \{predictions\}
    \item Evidence-gap report: \{evidence\_gap\_report\}
    \item Questions already asked: \{asked\_questions\}
    \item Interaction so far: \{dialog / summary\}
\end{itemize}

\textbf{Output format.}
\begin{verbatim}
{
  "question": "next diagnostic question"
}
\end{verbatim}

\end{HighlighterBox} \vspace{0.5cm}

\subsection{Doctor: Diagnosis Agents}
\label{appendix_prompts_diagnosis}

After each dialog turn, a \textbf{Diagnosis Agent} produces ranked set of top-3 differential diagnosis list with supporting reasoning and calibrated confidence scores. Several versions of this agent are used depending on the system configuration.
In the Specialist-Ensemble variant, 9 \textbf{Specialist agents} (fields Cardiology, Pulmonology, Gastroenterology, Neurology, Endocrinology, Nephrology, Rheumatology, Hematology, Dermatology) along with one generalist \textbf{Diagnosis Agent} independently produce top-3 differentials from the clinical information collected so far; a \textbf{Selector-SE Agent} then consolidates these 30 candidate predictions into a single top-3 differential set with reasoning and confidence scores, weighting each specialist’s opinion by its relevance to the case.
In the Knowledge-Base variant, the top-3 diagnoses derived from the UBKG-retrieved candidate diseases are combined with the predictions of the general \textbf{Diagnosis Agent} using a \textbf{Selector-KG Agent}, which produces a final top-3 differential set with reasoning and confidence scores.
In the \medbot variant, the output of the \textbf{Selector-SE} agent and top-3 UBKG candidates is filtered down to three diagnosis using the \textbf{Selector} agent.

\begin{HighlighterBox}{Diagnosis Agent Prompt}

\textbf{Role.}
You are an experienced diagnostician analyzing a doctor-patient consultation.

\textbf{Instruction.}
Predict the top-3 most likely diagnoses based on the available dialogue.

\textbf{Rules.}
\begin{itemize}
    \item Use all information provided in the dialogue.
    \item Provide exactly three distinct diagnoses, ranked from most to least likely.
    \item Be as specific as possible, including disease subtype when supported.
    \item Assign calibrated confidence scores from 0 to 100.
    \item Do not output ``healthy'', ``no illness'', or similar non-diagnoses.
    \item If information is insufficient, provide the best differential diagnoses with low confidence.
\end{itemize}

\textbf{Confidence calibration.}
\begin{itemize}
    \item 95--100: definitive or pathognomonic evidence.
    \item 80--90: strong clinical picture with supportive objective evidence.
    \item 70--80: likely diagnosis with remaining uncertainty.
    \item 50--69: possible diagnosis requiring more information.
    \item $<$50: weak or uncertain diagnosis.
\end{itemize}

\textbf{Input.}
\begin{itemize}
    \item Interaction so far: \{dialog / summary\}
\end{itemize}

\textbf{Output format.}
\begin{verbatim}
"predictions": 
[{"disease": ".", "conf": 85, "reason": "."},
{"disease": ".", "conf": 70, "reason": "."},
{"disease": ".", "conf": 55, "reason": "."}]
\end{verbatim}

\end{HighlighterBox} \vspace{0.5cm}

\begin{HighlighterBox}{Specialist Agent Prompt (Ex. Cardiology)}

\textbf{Role.}
You are an experienced cardiologist analyzing a doctor-patient consultation.

\textbf{Instruction.}
Predict the top-3 most likely cardiovascular diagnoses based on the available dialogue.

\textbf{Rules.}
\begin{itemize}
    \item Use all information provided in the dialogue.
    \item Prioritize cardiac and vascular causes when supported by the evidence.
    \item Provide exactly three distinct diagnoses, ranked from most to least likely.
    \item Be as specific as possible, including disease subtype when supported.
    \item Assign calibrated confidence scores from 0 to 100.
    \item Do not output ``healthy'', ``no illness'', or similar non-diagnoses.
    \item If information is insufficient, provide the best differential diagnoses with low confidence.
\end{itemize}

\textbf{Cardiology-specific considerations.}
\begin{itemize}
    \item Consider ischemic, heart failure, arrhythmic, valvular, pericardial, and vascular conditions.
    \item Evaluate chest pain, dyspnea, palpitations, syncope, edema, and cardiovascular risk factors.
    \item Use relevant investigations such as ECG, cardiac biomarkers, echocardiography, and stress testing when available.
    \item Distinguish cardiac from pulmonary or non-cardiac causes where appropriate.
    \item Consider cardiovascular emergencies including acute coronary syndrome, heart failure exacerbation, pulmonary embolism, and aortic dissection.
\end{itemize}

\textbf{Confidence calibration.}
\begin{itemize}
    \item 95--100: definitive or pathognomonic evidence.
    \item 80--90: strong clinical picture with supportive objective evidence.
    \item 70--80: likely diagnosis with remaining uncertainty.
    \item 50--69: possible diagnosis requiring more information.
    \item $<$50: weak or uncertain diagnosis.
\end{itemize}

\textbf{Input.}
\begin{itemize}
    \item Interaction so far: \{dialog / summary\}
\end{itemize}

\textbf{Output format.}
\begin{verbatim}
"predictions": 
[{"disease": ".", "conf": 85, "reason": "."},
{"disease": ".", "conf": 70, "reason": "."},
{"disease": ".", "conf": 55, "reason": "."}]
\end{verbatim}

\end{HighlighterBox} \vspace{0.5cm}

\begin{HighlighterBox}{Selector-SE Agent Prompt}

\textbf{Role.}
You are a senior attending physician reviewing diagnoses from 10 physicians who independently examined the same patient: 9 specialists (Cardiology, Pulmonology, Gastroenterology, Neurology, Endocrinology, Nephrology, Rheumatology, Hematology, Dermatology) and 1 general physician.

\textbf{Instruction.}
Each physician provided their top-3 diagnoses with confidence scores and reasoning (30 candidates total). Use the doctor--patient dialog as primary evidence to determine which specialists are relevant to the case and weigh their opinions accordingly.

\textbf{Rules.}
\begin{itemize}
    \item Do not perform simple majority voting. A diagnosis from the single relevant specialist outweighs five irrelevant ones.
    \item Give more weight to specialists whose domain matches the clinical presentation; ignore or downweight clearly irrelevant domains.
    \item Always consider the general physician's opinion as an unbiased baseline.
    \item Cross-domain diagnoses from unrelated specialists may still be valid if supported by dialog evidence.
    \item Be specific with disease subtypes when the evidence supports it.
    \item Provide exactly three diagnoses with calibrated confidence.
\end{itemize}

\textbf{Input.}
\begin{itemize}
    \item Interaction so far: \{dialog / summary\}
    \item Specialist opinions (10 $\times$ top-3): \{specialist\_predictions\}
\end{itemize}

\textbf{Output format.}
\begin{verbatim}
"predictions": 
[{"disease": ".", "conf": 85, "reason": "."},
{"disease": ".", "conf": 70, "reason": "."},
{"disease": ".", "conf": 55, "reason": "."}]
\end{verbatim}

\end{HighlighterBox} \vspace{0.5cm}

\begin{HighlighterBox}{Selector-KG Agent Prompt}

\textbf{Role.}
You are a senior diagnostician making a final diagnostic decision from multiple diagnostic sources.

\textbf{Instruction.}
Use the clinical summary as primary evidence, while considering diagnoses proposed by the knowledge graph and the LLM diagnostician. Produce an independent final top-3 diagnosis list.

\textbf{Rules.}
\begin{itemize}
    \item Do not simply merge or vote between the two lists.
    \item Evaluate the reasoning behind each proposed diagnosis.
    \item Trust the clinical evidence over source agreement.
    \item Prioritize objective data, including laboratory results, imaging, physical examination, and vital signs.
    \item Exclude diagnoses contradicted by absent symptoms or objective findings.
    \item Provide exactly three diagnoses with calibrated confidence.
\end{itemize}

\textbf{Input.}
\begin{itemize}
    \item Interaction so far: \{dialog / summary\}
    \item Knowledge-base diagnoses: \{kb\_preds\}
    \item LLM diagnoses: \{llm\_predictions\}
\end{itemize}

\textbf{Output format.}
\begin{verbatim}
"predictions": 
[{"disease": ".", "conf": 85, "reason": "."},
{"disease": ".", "conf": 70, "reason": "."},
{"disease": ".", "conf": 55, "reason": "."}]
\end{verbatim}

\end{HighlighterBox} \vspace{0.5cm}

\begin{HighlighterBox}{Selector Agent Prompt}

\textbf{Role.}
You are a senior diagnostician making a final diagnostic decision from multiple diagnostic sources.

\textbf{Instruction.}
Use the clinical summary as primary evidence, while considering candidate diagnoses from the knowledge graph and the specialist-ensemble predictions. Produce an independent final top-3 diagnosis list.

\textbf{Rules.}
\begin{itemize}
    \item Do not simply merge or vote between the two lists.
    \item Evaluate the reasoning behind each proposed diagnosis.
    \item Trust the clinical evidence over source agreement.
    \item Prioritize objective data, including laboratory results, imaging, physical examination, and vital signs.
    \item Exclude diagnoses contradicted by absent symptoms or objective findings.
    \item Provide exactly three diagnoses with calibrated confidence.
\end{itemize}

\textbf{Input.}
\begin{itemize}
    \item Interaction so far: \{dialog / summary\}
    \item Knowledge-base diagnoses: \{kb\_preds\}
    \item Specialist ensemble predictions: \{specialist\_predictions\}
\end{itemize}

\textbf{Output format.}
\begin{verbatim}
"predictions": 
[{"disease": ".", "conf": 85, "reason": "."},
{"disease": ".", "conf": 70, "reason": "."},
{"disease": ".", "conf": 55, "reason": "."}]
\end{verbatim}

\end{HighlighterBox} \vspace{0.5cm}

\subsection{Summarizer Agents}
\label{appendix_prompts_summarizer}
The Summarizer distills a multi-turn dialogue encounter into a concise case-vignette style representation. We evaluate two formats: a free-form \emph{Summarizer: Paragraph} and a \emph{Summarizer: Structured} that fills predefined clinical fields. In both the Summarizer variant and the combined \medbot{} variant, summaries are provided to all downstream agents in place of the full dialogue history.

\begin{HighlighterBox}{Summarizer: Paragraph Prompt}

\textbf{Role.}
You are a medical scribe summarizing a doctor-patient consultation.

\textbf{Instruction.}
Create a concise case summary from the dialogue for downstream diagnosis.

\textbf{Rules.}
\begin{itemize}
    \item Include only information explicitly stated by the patient.
    \item Do not infer or add information.
    \item Organize the summary logically by demographics, chief complaint, symptoms, history, examination findings, and test results.
    \item Use appropriate medical terminology.
    \item Write in third person.
    \item Keep the summary concise but complete.
\end{itemize}

\textbf{Output.}
Return only the case summary, without headers or explanations.

\end{HighlighterBox} \vspace{0.5cm}

\begin{HighlighterBox}{Summarizer: Structured Prompt}

\textbf{Role.}
You are a medical scribe maintaining a structured diagnostic state from a doctor-patient consultation.

\textbf{Instruction.}
Update the structured clinical summary using only newly stated information from the latest dialogue turn.

\textbf{Structured summary fields.}
\begin{verbatim}
{
  "demographics": "",
  "vital_signs": [],
  "symptoms_present": [],
  "symptoms_absent": [],
  "symptom_details": [],
  "onset": "",
  "medical_history": [],
  "medications": [],
  "physical_exam": [],
  "risk_factors": [],
  "recent_exposures": [],
  "tests": []
}
\end{verbatim}

\textbf{Rules.}
\begin{itemize}
    \item Include only explicitly stated patient information.
    \item Do not infer or assume unstated facts.
    \item Add new findings without removing existing ones.
    \item Avoid duplicate entries.
    \item Put denied symptoms in \texttt{symptoms\_absent}.
    \item Put laboratory, imaging, ECG, biopsy, culture, and other diagnostic results in \texttt{tests}.
    \item Use simple canonical symptom names in \texttt{symptoms\_present} and \texttt{symptoms\_absent}; place severity, timing, location, triggers, and relieving factors in \texttt{symptom\_details}.
    \item Preserve numerical values when provided.
\end{itemize}

\textbf{Output.}
Return the updated clinical summary in the exact JSON format above.

\end{HighlighterBox} \vspace{0.5cm}

\subsection{Evidence Gap Agent}
\label{appendix_prompts_evidence}

The Evidence Gap agent identifies clinically informative missing evidence from the current case state. In early turns, it highlights general missing information (e.g., vitals, labs, history details), while after differential questioning begins, it additionally identifies diagnosis-specific gaps that would help distinguish among the current top candidate diagnoses. These gaps are then used by the Evidence-Guided question agents (EG-QA and EG-DQA, Section~\ref{appendix_prompts_question}) to guide subsequent question selection.

\begin{HighlighterBox}{Evidence Gap Agent Prompt}

\textbf{Role.}
You are a clinical evidence analyst identifying missing diagnostic information.

\textbf{Instruction.}
Given the current clinical summary and top-3 differential diagnoses, produce an evidence-gap report.

\textbf{Gap types.}
\begin{itemize}
    \item \textbf{General gaps:} missing baseline information useful for any diagnostic workup, such as demographics, vitals, symptom timeline, medical history, medications, family history, exposures, and basic tests.
    \item \textbf{Diagnosis-specific gaps:} missing evidence required to confirm or rule out each candidate diagnosis.
\end{itemize}

\textbf{Rules.}
\begin{itemize}
    \item Focus on actionable information that can be obtained from the patient or standard tests.
    \item Prioritize gaps that would most strongly shift confidence between competing diagnoses.
    \item Include red-flag gaps when missing information could indicate serious disease.
    \item Provide concise, clinically useful missing-evidence statements.
\end{itemize}

\textbf{Input.}
\begin{itemize}
    \item Clinical summary: \{summary\}
    \item Current top-3 differential diagnoses: \{predictions\}
\end{itemize}

\textbf{Output format.}
\begin{verbatim}
{
  "general_gaps": [
    "missing general information item"
  ],
  "diagnosis_specific_gaps": [
    {
      "disease": "Diagnosis name",
      "missing_evidence": "Required evidence"
    }
  ]
}
\end{verbatim}

\end{HighlighterBox} \vspace{0.5cm}

\subsection{Judge Agent}
\label{appendix_prompts_judge}

The Judge evaluates the final predicted diagnosis against the ground-truth label using semantic matching, accepting exact matches, synonyms, paraphrases and more specific subtypes of the ground-truth diagnosis (not vice versa), while rejecting different or even near-misses across distinct conditions (e.g., Type 1 vs.\ Type 2 diabetes).

\begin{HighlighterBox}{Judge Agent Prompt}

\textbf{Role.}
You are a medical terminology expert comparing disease names.

\textbf{Instruction.}
Determine whether the predicted disease and the ground-truth disease refer to the same medical condition.

\textbf{Rules.}
\begin{itemize}
    \item Treat medical synonyms, abbreviations, and common-name variants as equivalent.
    \item Treat a predicted disease as correct if it is a more specific subtype of the ground truth.
    \item Be strict when clinically distinct subtypes are involved; they should not be treated as the same condition, e.g., Type 1 vs. Type 2 diabetes.
    \item Return false when the diseases are related but not equivalent.
\end{itemize}

\textbf{Input.}
\begin{itemize}
    \item Predicted disease: \{predicted\}
    \item Ground-truth disease: \{ground\_truth\}
\end{itemize}

\textbf{Output format.}
\begin{verbatim}
{
  "same": true/false,
  "explanation": "brief explanation"
}
\end{verbatim}

\end{HighlighterBox} \vspace{0.5cm}
\section{Example}
\label{appendix_qualitative}

\begin{DialogueExampleBox}{Case ID: pubmed\_35800852}

\textbf{Specialty:} Gastroenterology.

\medskip

\textbf{Ground Truth:} IgG4-related hepatic inflammatory pseudotumor; IgG4-related autoimmune pancreatitis (Type I).

\medskip

\textbf{Case Description:}

\textit{A 59-year-old incarcerated male with human immunodeficiency virus on highly active antiretroviral therapy presented with abdominal pain and jaundice. His liver panel was notably abnormal, with alkaline phosphatase 344 U/L (25--100 U/L), ALT 61 U/L (10--40 U/L), AST 44 U/L (8--40 U/L), total bilirubin 11.3 mg/dL (0.3--1.2 mg/dL), and direct bilirubin 7.2 mg/dL (0.0--0.4 mg/dL). An initial computed tomography (CT) scan revealed an amorphous-appearing pancreatic head mass with dilated intra- and extrahepatic bile ducts. A serum IgG4 level measured 354 mg/dL (4--86 mg/dL), and cancer antigen (CA) 19--9 was 97.9 U/ml (0--35.0 U/mL). Endoscopic ultrasound (EUS)-guided fine needle biopsy of the pancreatic head mass was performed. Pathology was consistent with type I autoimmune pancreatitis by immunohistochemistry, and no malignant cells were identified. Due to significant biliary obstruction, he underwent an endoscopic retrograde cholangiopancreatography (ERCP) with stenting of the distal inflammatory biliary stricture and initiated prednisone therapy with an extended taper, resulting in subsequent normalization of CT findings and liver labs.}

\smallskip

\textit{Two years later, he again presented with similar symptoms. His liver enzymes were again elevated in a cholestatic pattern with a total bilirubin of 4.0 mg/dL, and his serum IgG4 was 458.9 mg/dL. The CT scan showed a mass-like mural thickening at the hepatic hilum concerning for a Klatskin tumor. EUS confirmed an irregular hypoechoic mass in the porta hepatis measuring 4.1 $\times$ 2.7 cm. Histopathology showed dense lymphoplasmacytic infiltrating and a fibrosing process with residual bland pancreatic ducts and lobules. Immunohistochemistry demonstrated IgG4 plasma cell positivity ($>$10 per high-power field (HPF)). A subsequent CT-guided biopsy of the hilar mass also confirmed IgG4-positive plasma cells and obliterative phlebitis, consistent with IgG4-associated hepatic IPT. He was again treated with a tapered course of prednisone with repeat imaging showing complete resolution of the pseudotumor.}

\vspace{4mm}\hrule\vspace{4mm}

\textbf{Turn 1}

\medskip

\textbf{Doctor:} What is your age, gender, and occupation?

\medskip

\textbf{Patient:} I am a 59-year-old male, and I am incarcerated.

\medskip

\textit{Paragraph Summary:}
The patient is a 59-year-old male who is currently incarcerated.

\medskip

\textit{Diagnosis (top 3):}
\begin{enumerate}\setlength\itemsep{2pt}
  \item Chronic Obstructive Pulmonary Disease (COPD) (55\%) -- age and potential exposure to smoking or environmental pollutants in a confined setting.
  \item Hypertension (50\%) -- common in older males, especially in stressful environments like incarceration.
  \item Type 2 Diabetes Mellitus (50\%) -- age and lifestyle factors associated with incarceration.
\end{enumerate}



\vspace{4mm}\hrule\vspace{4mm}

\textbf{Turn 2}

\medskip

\textbf{Doctor:} What is your chief complaint or main symptom that brought you to seek medical attention?

\medskip

\textbf{Patient:} I have abdominal pain and jaundice.

\medskip

\textit{Paragraph Summary:}
The patient is a 59-year-old male who is incarcerated. He reports abdominal pain and jaundice as his chief complaints.

\medskip

\textit{Diagnosis (top 3):}
\begin{enumerate}\setlength\itemsep{2pt}
  \item Acute viral hepatitis (e.g., Hepatitis B or C) (85\%) -- hallmark abdominal pain + jaundice; incarcerated patient with bloodborne-pathogen exposure risk.
  \item Cholelithiasis with choledocholithiasis (70\%) -- bile duct obstruction from gallstones in an older male.
  \item Pancreatic cancer (55\%) -- jaundice + abdominal pain from possible bile duct obstruction.
\end{enumerate}

\medskip

\textit{Evidence Gap:}
Vital signs; pain characterisation; associated symptoms (nausea, vomiting, weight loss, fever, bowel habits); past medical history; current medications; allergies; family/social history; physical exam; basic labs and imaging.

\vspace{4mm}\hrule\vspace{4mm}

\textbf{Turn 3}

\medskip

\textbf{Doctor:} Can you describe the abdominal pain in detail, including its onset, duration and severity?

\medskip

\textbf{Patient:} I have abdominal pain, but I don't know the specific details about its onset, duration, or severity.

\vspace{4mm}\hrule\vspace{4mm}

\textbf{Turn 4}

\medskip

\textbf{Doctor:} Do you have any associated symptoms such as nausea, vomiting, weight loss, fever, or changes in bowel habits?

\medskip

\textbf{Patient:} I don't know.

\vspace{4mm}\hrule\vspace{4mm}

\textbf{Turn 5}

\medskip

\textbf{Doctor:} What are your current vital signs, including blood pressure, heart rate, temperature, oxygen saturation, and respiratory rate?

\medskip

\textbf{Patient:} I don't know.

\vspace{4mm}\hrule\vspace{4mm}

\textbf{Turn 6}

\medskip

\textbf{Doctor:} Can you provide a detailed medical history, including any past illnesses, chronic conditions, surgeries, or hospitalizations?

\medskip

\textbf{Patient:} I have a history of human immunodeficiency virus and have been treated for autoimmune pancreatitis and IgG4-associated hepatic inflammatory pseudotumor. I have undergone procedures including endoscopic retrograde cholangiopancreatography with stenting and endoscopic ultrasound-guided fine needle biopsy. I have been treated with prednisone therapy for these conditions.

\medskip

\textit{Paragraph Summary:}
The patient is a 59-year-old male who is incarcerated. He presents with abdominal pain and jaundice. He is unable to provide specific details about the abdominal pain or associated symptoms. His medical history includes human immunodeficiency virus, autoimmune pancreatitis, and IgG4-associated hepatic inflammatory pseudotumor. He has undergone ERCP with stenting and EUS-guided fine needle biopsy, and has been treated with prednisone therapy.

\medskip

\textit{Diagnosis (top 3):}
\begin{enumerate}\setlength\itemsep{2pt}
  \item IgG4-related sclerosing cholangitis (85\%) -- history of IgG4-associated hepatic IPT and autoimmune pancreatitis combined with jaundice and abdominal pain.
  \item Pancreatic cancer (70\%) -- history of autoimmune pancreatitis raises concern; chronic inflammation a risk factor.
  \item HIV-associated cholangiopathy (55\%) -- HIV status with jaundice; less likely than IgG4-related disease given specific history.
\end{enumerate}

\medskip

\textit{Evidence Gap:}
Detailed vital signs; specific abdominal pain characterisation; associated symptoms; current medication dosing; allergies; family/social history; physical exam; basic labs and imaging.

\vspace{4mm}\hrule\vspace{4mm}

\textbf{Turn 7}

\medskip

\textbf{Doctor:} Do you have any known allergies, including drug, food, or environmental allergies?

\medskip

\textbf{Patient:} I don't know.

\vspace{4mm}\hrule\vspace{4mm}

\textbf{Turn 8}

\medskip

\textbf{Doctor:} What medications are you currently taking, including dosages, frequency, and any recent changes?

\medskip

\textbf{Patient:} I am currently on prednisone therapy with a tapered dosage schedule, but I don't know the exact dosage or frequency.

\medskip

\textit{Paragraph Summary:}
The patient is a 59-year-old male who is incarcerated. He presents with abdominal pain and jaundice. He is unable to provide specific details about the abdominal pain or associated symptoms. His medical history includes human immunodeficiency virus, autoimmune pancreatitis, and IgG4-associated hepatic inflammatory pseudotumor. He has undergone procedures such as ERCP with stenting and EUS-guided fine needle biopsy. He has been treated with prednisone therapy and is currently on a tapered dosage schedule, though the exact dosage and frequency are unknown. The patient is unaware of his current vital signs and any allergies.

\medskip

\textit{Diagnosis (top 3):}
\begin{enumerate}\setlength\itemsep{2pt}
  \item IgG4-related sclerosing cholangitis (85\%) -- history of IgG4-associated hepatic IPT and autoimmune pancreatitis strongly associated with IgG4-RSC; jaundice and abdominal pain align.
  \item Pancreatic cancer (70\%) -- history of autoimmune pancreatitis and jaundice raise concern for pancreatic malignancy; chronic pancreatitis is a risk factor.
  \item Chronic hepatitis with cirrhosis (55\%) -- history of hepatic pseudotumor and jaundice could indicate underlying chronic liver disease; less likely but worth considering.
\end{enumerate}

\medskip

\textit{Evidence Gap:}
Exact current vital signs; detailed abdominal pain characterisation; associated symptoms (nausea, vomiting, weight loss, fever, bowel changes); complete medication list with dosing; allergies; family history; social history (smoking, alcohol, drug use); review of systems; physical exam; basic labs (LFTs, CBC) and imaging.

\vspace{4mm}\hrule\vspace{4mm}

\textbf{Turn 9}

\medskip

\textbf{Doctor:} Do you have a family history of any hereditary conditions or diseases, such as liver disease, autoimmune disorders, or cancer?

\medskip

\textbf{Patient:} I don't know.

\vspace{4mm}\hrule\vspace{4mm}

\textbf{Turn 10}

\medskip

\textbf{Doctor:} Can you provide details about your social history, including smoking, alcohol use, or drug use?

\medskip

\textbf{Patient:} I don't know.

\vspace{4mm}\hrule\vspace{4mm}

\textbf{Turn 11}

\medskip

\textbf{Doctor:} Have you undergone any recent laboratory tests, such as liver function tests, complete blood count, or imaging studies, and can you provide the results?

\medskip

\textbf{Patient:} Yes, my recent tests showed elevated liver enzymes in a cholestatic pattern, a total bilirubin of 4.0 mg/dL, and a serum IgG4 level of 458.9 mg/dL. Imaging revealed a mass-like mural thickening at the hepatic hilum, and further biopsy confirmed IgG4-positive plasma cells and obliterative phlebitis.

\medskip

\textit{Paragraph Summary:}
The patient is a 59-year-old male who is incarcerated, presenting with abdominal pain and jaundice. He has a medical history of human immunodeficiency virus, autoimmune pancreatitis, and IgG4-associated hepatic inflammatory pseudotumor, with prior treatments including endoscopic retrograde cholangiopancreatography with stenting, endoscopic ultrasound-guided fine needle biopsy, and prednisone therapy. Current medications include prednisone on a tapered dosage schedule. Recent laboratory tests revealed elevated liver enzymes in a cholestatic pattern, a total bilirubin of 4.0 mg/dL, and a serum IgG4 level of 458.9 mg/dL. Imaging showed a mass-like mural thickening at the hepatic hilum, and biopsy confirmed IgG4-positive plasma cells and obliterative phlebitis.

\medskip

\textit{Specialist Ensemble (top 3):}
\begin{enumerate}\setlength\itemsep{2pt}
  \item IgG4-related sclerosing cholangitis (95\%) -- jaundice, cholestatic LFTs, elevated IgG4, mural thickening at hepatic hilum, biopsy with IgG4+ plasma cells and obliterative phlebitis; multiple specialists agree.
  \item Autoimmune pancreatitis (80\%) -- history + elevated IgG4 suggest recurrence or associated involvement.
  \item Cholangiocarcinoma (65\%) -- mass-like hilar thickening; malignancy cannot be fully excluded without further investigation.
\end{enumerate}

\medskip



\textit{KB Diagnosis (top 3):}
\begin{enumerate}\setlength\itemsep{2pt}
  \item IgG4-associated cholangitis
  \item Autoimmune pancreatitis
  \item \textbf{IgG4-associated hepatic inflammatory pseudotumor}
\end{enumerate}

\medskip

\textit{Selector (top 3):}
\begin{enumerate}\setlength\itemsep{2pt}
  \item IgG4-related sclerosing cholangitis (90\%) -- jaundice, cholestatic LFTs, elevated IgG4, mural thickening, biopsy with IgG4+ plasma cells and obliterative phlebitis.
  \item Autoimmune pancreatitis (75\%) -- history + elevated IgG4 suggest recurrence; primary findings still favor cholangitis.
  \item Cholangiocarcinoma (60\%) -- mass-like hilar thickening; biopsy favors IgG4-RD but malignancy not fully excluded.
\end{enumerate}

\medskip

\textit{General Evidence Gap:}
Detailed vital signs; symptom timeline (onset, progression, triggers, relieving factors); family history of autoimmune or hepatic conditions; recent changes/adherence to the prednisone taper.

\medskip

\textit{Diagnosis-Specific Evidence Gap:}
\begin{itemize}\setlength\itemsep{2pt}
  \item \textit{IgG4-related sclerosing cholangitis:} MRCP for bile-duct visualization; additional autoimmune serological markers; follow-up biopsy focused on IgG4+ plasma cell infiltration and fibrosis.
  \item \textit{Autoimmune pancreatitis:} contrast-enhanced CT or MRI of the pancreas; serum amylase and lipase; review of symptom correlation with pancreatic involvement.
  \item \textit{Cholangiocarcinoma:} PET-CT for metabolic activity; tumor markers (CA 19-9); histopathological review focused on malignant features.
\end{itemize}

\vspace{4mm}\hrule\vspace{4mm}

\textbf{Turn 12}

\medskip

\textbf{Doctor:} Have you undergone advanced imaging studies such as magnetic resonance cholangiopancreatography (MRCP) or PET-CT recently, and if so, what were the findings?

\medskip

\textbf{Patient:} I don't know.

\vspace{4mm}\hrule\vspace{4mm}

\textbf{Turn 13}

\medskip

\textbf{Doctor:} Have serum tumor markers such as CA 19-9 been evaluated recently, and if so, what were the results?

\medskip

\textbf{Patient:} I don't know.

\vspace{4mm}\hrule\vspace{4mm}

\textbf{Turn 14}

\medskip

\textbf{Doctor:} Have serum markers specific to pancreatic inflammation, such as amylase and lipase levels, been evaluated recently, and if so, what were the results?

\medskip

\textbf{Patient:} I don't know.

\vspace{4mm}\hrule\vspace{4mm}

\textbf{Turn 15}

\medskip

\textbf{Doctor:} Have you recently undergone a follow-up biopsy of the hepatic hilum mass, and if so, what were the histopathological findings, particularly regarding IgG4-positive plasma cell infiltration and any evidence of malignancy?

\medskip

\textbf{Patient:} Yes, I recently underwent a follow-up biopsy of the hepatic hilum mass, and the histopathology showed dense lymphoplasmacytic infiltration, a fibrosing process, IgG4-positive plasma cells, and obliterative phlebitis, with no evidence of malignancy.

\medskip

\textit{Paragraph Summary:}
The patient is a 59-year-old male who is incarcerated. He presents with abdominal pain and jaundice. He has a medical history of human immunodeficiency virus, autoimmune pancreatitis, and IgG4-associated hepatic inflammatory pseudotumor, previously treated with prednisone therapy and procedures including ERCP with stenting and EUS-guided fine needle biopsy. Recent laboratory tests revealed elevated liver enzymes in a cholestatic pattern, total bilirubin of 4.0 mg/dL, and serum IgG4 level of 458.9 mg/dL. Imaging showed a mass-like mural thickening at the hepatic hilum, and biopsy confirmed IgG4-positive plasma cells and obliterative phlebitis. A follow-up biopsy of the hepatic hilum mass revealed dense lymphoplasmacytic infiltration, fibrosing process, IgG4-positive plasma cells, and obliterative phlebitis, with no evidence of malignancy. The patient is currently on a tapered prednisone therapy schedule.

\medskip

\textit{Specialist Ensemble (top 3):}
\begin{enumerate}\setlength\itemsep{2pt}
  \item IgG4-related sclerosing cholangitis (95\%) -- jaundice, cholestatic LFTs, elevated IgG4, biopsy with IgG4+ plasma cells and obliterative phlebitis; agreed by General Physician and Gastroenterology.
  \item Autoimmune pancreatitis (85\%) -- history + elevated IgG4 + systemic involvement; supported by General Physician, Gastroenterology, Rheumatology.
  \item \textbf{IgG4-related hepatic inflammatory pseudotumor} (80\%) -- mass-like hilar lesion + biopsy; aligns with patient history and IgG4-RD spectrum.
\end{enumerate}

\medskip

\textit{KB Diagnosis (top 3):}
\begin{enumerate}\setlength\itemsep{2pt}
  \item IgG4-associated cholangitis
  \item Autoimmune pancreatitis
  \item \textbf{IgG4-related hepatic inflammatory pseudotumor}
\end{enumerate}

\medskip

\textit{Selector (top 3):}
\begin{enumerate}\setlength\itemsep{2pt}
  \item IgG4-related sclerosing cholangitis (90\%) -- elevated IgG4, cholestatic LFTs, jaundice, biopsy with IgG4+ plasma cells and obliterative phlebitis.
  \item \textbf{IgG4-related hepatic inflammatory pseudotumor} (80\%) -- mass-like hilar lesion + dense lymphoplasmacytic infiltration + IgG4+ plasma cells.
  \item Autoimmune pancreatitis (70\%) -- history + elevated IgG4 + systemic involvement as part of IgG4-RD spectrum.
\end{enumerate}

\medskip

\textit{General Evidence Gap:}
Vital signs; symptom timeline (onset, exacerbating/relieving factors); exact prednisone dosing and other medications; family history of autoimmune or IgG4-related diseases.

\medskip

\textit{Diagnosis-Specific Evidence Gap:}
\begin{itemize}\setlength\itemsep{2pt}
  \item \textit{IgG4-related sclerosing cholangitis:} MRCP/CT of bile ducts; ALP and GGT to support cholestatic pattern; repeat biopsy focused on bile-duct involvement.
  \item \textit{IgG4-related hepatic inflammatory pseudotumor:} contrast-enhanced MRI of the hepatic hilum mass; IHC staining for IgG4 and related markers; monitoring response to prednisone therapy.
  \item \textit{Autoimmune pancreatitis:} pancreatic MRI or CT for structural changes/inflammation; serum amylase and lipase; history of recurrent pancreatitis episodes.
\end{itemize}

\vspace{4mm}\hrule\vspace{4mm}

\textbf{Turn 16}

\medskip

\textbf{Doctor:} Have you recently undergone imaging studies such as contrast-enhanced MRI or CT focusing on the hepatic hilum mass, and if so, what were the findings?

\medskip

\textbf{Patient:} Yes, I recently had a CT scan that showed a mass-like mural thickening at the hepatic hilum concerning for a Klatskin tumor.

\medskip

\textit{Paragraph Summary:}
The patient is a 59-year-old male who is incarcerated and presents with abdominal pain and jaundice. He has a medical history of human immunodeficiency virus, autoimmune pancreatitis, and IgG4-associated hepatic inflammatory pseudotumor, treated with prednisone therapy and procedures including ERCP with stenting and EUS-guided fine needle biopsy. Recent laboratory tests revealed elevated liver enzymes in a cholestatic pattern, total bilirubin of 4.0 mg/dL, and serum IgG4 level of 458.9 mg/dL. Imaging studies showed a mass-like mural thickening at the hepatic hilum, and biopsy confirmed IgG4-positive plasma cells, obliterative phlebitis, and dense lymphoplasmacytic infiltration with no evidence of malignancy. A recent CT scan raised concern for a Klatskin tumor.

\medskip

\textit{Specialist Ensemble (top 3):}
\begin{enumerate}\setlength\itemsep{2pt}
  \item IgG4-related sclerosing cholangitis (95\%) -- elevated IgG4, cholestatic LFTs, biopsy with IgG4+ plasma cells and obliterative phlebitis; multiple specialists agree.
  \item Klatskin tumor (cholangiocarcinoma) (70\%) -- CT findings of mass-like mural thickening at hepatic hilum; biopsy did not confirm malignancy, but cannot fully exclude.
  \item Autoimmune pancreatitis-associated cholangitis (60\%) -- history of autoimmune pancreatitis and IgG4-related disease.
\end{enumerate}

\medskip

\textit{KB Diagnosis (top 3):}
\begin{enumerate}\setlength\itemsep{2pt}
  \item IgG4-related sclerosing cholangitis
  \item Autoimmune pancreatitis
  \item Klatskin tumor (hilar cholangiocarcinoma)
\end{enumerate}

\medskip

\textit{Selector (top 3):}
\begin{enumerate}\setlength\itemsep{2pt}
  \item IgG4-related sclerosing cholangitis (90\%) -- elevated IgG4, IgG4+ plasma cells, obliterative phlebitis, mural thickening at hepatic hilum.
  \item Klatskin tumor (hilar cholangiocarcinoma) (65\%) -- CT concern for malignancy; biopsy negative.
  \item Autoimmune pancreatitis-associated cholangitis (55\%) -- history + IgG4-RD; evidence more strongly supports primary cholangitis.
\end{enumerate}

\medskip

\textit{General Evidence Gap:}
Vital signs; symptom timeline and triggers for abdominal pain and jaundice; current medication adherence and recent changes; family history of liver, pancreatic, autoimmune, or malignant disease.

\medskip

\textit{Diagnosis-Specific Evidence Gap:}
\begin{itemize}\setlength\itemsep{2pt}
  \item \textit{IgG4-related sclerosing cholangitis:} MRCP or PET to delineate biliary involvement and exclude malignancy; serial serum IgG4 + corticosteroid response; biopsy for storiform fibrosis.
  \item \textit{Klatskin tumor (hilar cholangiocarcinoma):} repeat biopsy with molecular markers/IHC; tumor markers (CA 19-9, CEA); contrast-enhanced MRI or CT angiography for vascular invasion/metastasis.
  \item \textit{Autoimmune pancreatitis-associated cholangitis:} serum amylase and lipase; pancreatic MRI or EUS; response to corticosteroid therapy for pancreatic symptoms.
\end{itemize}

\vspace{4mm}\hrule\vspace{4mm}

\textbf{Turn 17}

\medskip

\textbf{Doctor:} Have you undergone pancreatic imaging studies, such as MRI or endoscopic ultrasound (EUS), recently, and if so, what were the findings regarding pancreatic involvement?

\medskip

\textbf{Patient:} Yes, I underwent an endoscopic ultrasound (EUS) recently, which confirmed an irregular hypoechoic mass in the porta hepatis measuring 4.1 $\times$ 2.7 cm, consistent with IgG4-associated hepatic IPT.

\medskip

\textit{Paragraph Summary:}
The patient is a 59-year-old male who is incarcerated, presenting with abdominal pain and jaundice. He has a medical history of human immunodeficiency virus, autoimmune pancreatitis, and IgG4-associated hepatic inflammatory pseudotumor, previously treated with prednisone therapy and procedures such as ERCP with stenting and EUS-guided fine needle biopsy. Recent laboratory tests revealed elevated liver enzymes in a cholestatic pattern, total bilirubin of 4.0 mg/dL, and serum IgG4 level of 458.9 mg/dL. Imaging studies showed a mass-like mural thickening at the hepatic hilum, confirmed by biopsy to have dense lymphoplasmacytic infiltration, fibrosing process, IgG4-positive plasma cells, and obliterative phlebitis, with no evidence of malignancy. A CT scan suggested a Klatskin tumor, and EUS identified an irregular hypoechoic mass in the porta hepatis measuring 4.1 $\times$ 2.7 cm, consistent with IgG4-associated hepatic IPT.

\medskip

\textit{Specialist Ensemble (top 3):}
\begin{enumerate}\setlength\itemsep{2pt}
  \item IgG4-related sclerosing cholangitis (95\%) -- jaundice, cholestatic LFTs, elevated IgG4, biopsy with IgG4+ plasma cells and obliterative phlebitis; agreed by General Physician, Gastroenterology, Rheumatology.
  \item \textbf{IgG4-associated hepatic inflammatory pseudotumor} (85\%) -- mass-like hilar lesion + biopsy + elevated IgG4; agreed by General Physician and Gastroenterology.
  \item Autoimmune pancreatitis with hepatic involvement (75\%) -- history + current hepatic findings as systemic IgG4-RD manifestation.
\end{enumerate}

\medskip

\textit{KB Diagnosis (top 3):}
\begin{enumerate}\setlength\itemsep{2pt}
  \item \textbf{IgG4-associated hepatic inflammatory pseudotumor}
  \item Cholangiocarcinoma
  \item Autoimmune pancreatitis
\end{enumerate}

\medskip

\textit{Selector (top 3):}
\begin{enumerate}\setlength\itemsep{2pt}
  \item \textbf{IgG4-associated hepatic inflammatory pseudotumor (90\%)} -- elevated IgG4, IgG4+ plasma cells, mass-like hilar lesion; consistent with history of IgG4-related disease.
  \item IgG4-related sclerosing cholangitis (80\%) -- cholestatic LFT pattern, jaundice, obliterative phlebitis and IgG4+ plasma cells; known IgG4-RD manifestation.
  \item Cholangiocarcinoma (60\%) -- imaging concern for Klatskin tumor; biopsy negative for malignancy but lower-confidence differential.
\end{enumerate}

\medskip

\textit{General Evidence Gap:}
Vital signs; progression, duration, and triggers of abdominal pain and jaundice; current medications and adherence; family history of liver or autoimmune disease.

\medskip

\textit{Diagnosis-Specific Evidence Gap:}
\begin{itemize}\setlength\itemsep{2pt}
  \item \textit{IgG4-associated hepatic inflammatory pseudotumor:} contrast-enhanced MRI of the hepatic lesion; serial IgG4 levels and corticosteroid response; systemic evaluation for other IgG4-RD manifestations.
  \item \textit{IgG4-related sclerosing cholangitis:} direct cholangiography (MRCP/ERCP) for bile-duct strictures; bile-duct histology if accessible; assessment of other organ systems for IgG4-RD.
  \item \textit{Cholangiocarcinoma:} repeat biopsy with molecular markers/IHC; tumor markers (CA 19-9, CEA); PET-CT to assess metabolic activity and metastasis.
\end{itemize}

\vspace{4mm}\hrule\vspace{4mm}

\textbf{Turn 18}

\medskip

\textbf{Doctor:} Have you undergone direct cholangiography, such as MRCP or ERCP, recently, and if so, what were the findings regarding bile duct involvement or strictures?

\medskip

\textbf{Patient:} I don't know.

\vspace{4mm}\hrule\vspace{4mm}

\textbf{Turn 19}

\medskip

\textbf{Doctor:} Have serum tumor markers such as CA 19-9 and CEA been evaluated recently, and if so, what were the results?

\medskip

\textbf{Patient:} I don't know.

\vspace{4mm}\hrule\vspace{4mm}

\textbf{Turn 20}

\medskip

\textbf{Doctor:} Have you undergone direct cholangiography, such as ERCP or MRCP, recently, and if so, what were the findings regarding bile duct involvement or strictures?

\medskip

\textbf{Patient:} I don't know.

\vspace{4mm}\hrule\vspace{4mm}

\textbf{Final Prediction:} \textit{IgG4-associated hepatic inflammatory pseudotumor} -- judged \textbf{correct} against ground truth.

\end{DialogueExampleBox}
\end{document}